\documentclass[twocolumn]{aastex63}
\usepackage{enumerate}
\usepackage{amssymb, amsmath}
\usepackage{natbib}
\usepackage{color}
\usepackage{ulem}
\usepackage{appendix}
\usepackage{hyperref}
\usepackage[stable]{footmisc}
\usepackage{relsize}
\usepackage{mathtools}

\definecolor{red}{rgb}{1.0,0.0,0.0}

\begin{document}

\title{Dynamical and Atmospheric Characterization of the Substellar Companion HD 33632 Ab from Direct Imaging, Astrometry, and Radial-Velocity Data\footnote{Based in part on data collected at Subaru Telescope, which is operated by the National Astronomical Observatory of Japan.}}
\email{mona.elmorsy@utsa.edu}
\author{Mona El Morsy}
\affiliation{Department of Physics and Astronomy, University of Texas-San Antonio, San Antonio, TX, USA}
\author{Thayne Currie}
\affiliation{Department of Physics and Astronomy, University of Texas-San Antonio, San Antonio, TX, USA}
\affiliation{Subaru Telescope, National Astronomical Observatory of Japan, 
650 North A`oh$\bar{o}$k$\bar{u}$ Place, Hilo, HI  96720, USA}
\author{Danielle Bovie}
\affiliation{Department of Physics and Astronomy, University of Texas-San Antonio, San Antonio, TX, USA}
\author{Masayuki Kuzuhara}
\affiliation{Astrobiology Center of NINS, 2-21-1, Osawa, Mitaka, Tokyo, 181-8588, Japan}
\affiliation{National Astronomical Observatory of Japan, 2-21-2, Osawa, Mitaka, Tokyo 181-8588, Japan}
\author{Brianna Lacy}
\affiliation{Department of Astronomy and Astrophysics, University of California-Santa Cruz, Santa Cruz, CA, USA}
\author{Yiting Li}
\affiliation{Department of Astronomy, University of Michigan, 1085 S. University, Ann Arbor, MI 48109, USA}
\author{Taylor Tobin}
\affiliation{Department of Astronomy, University of Michigan, 1085 S. University, Ann Arbor, MI 48109, USA}
\author{Timothy D. Brandt}
\affiliation{Space Telescope Science Institute, Baltimore, MD, USA}
\affiliation{Department of Physics, University of California, Santa Barbara, Santa Barbara, California, USA}
\author{Jeffrey Chilcote}
\affiliation{Department of Physics, University of Notre Dame, South Bend, IN, USA}
\author{Olivier Guyon}
\affiliation{Subaru Telescope, National Astronomical Observatory of Japan, 
650 North A`oh$\bar{o}$k$\bar{u}$ Place, Hilo, HI  96720, USA}
\affil{Astrobiology Center of NINS, 2-21-1, Osawa, Mitaka, Tokyo, 181-8588, Japan}
\affil{Steward Observatory, The University of Arizona, Tucson, AZ 85721, USA}
\affil{College of Optical Sciences, University of Arizona, Tucson, AZ 85721, USA}
\author{Tyler D. Groff}
\affiliation{NASA-Goddard Space Flight Center, Greenbelt, MD, USA}
\author{Julien Lozi}
\affiliation{Subaru Telescope, National Astronomical Observatory of Japan, 
650 North A`oh$\bar{o}$k$\bar{u}$ Place, Hilo, HI  96720, USA}
\author{Sebastien Vievard}
\affiliation{Subaru Telescope, National Astronomical Observatory of Japan, 
650 North A`oh$\bar{o}$k$\bar{u}$ Place, Hilo, HI  96720, USA}
\author{Vincent Deo}
\affiliation{Subaru Telescope, National Astronomical Observatory of Japan, 
650 North A`oh$\bar{o}$k$\bar{u}$ Place, Hilo, HI  96720, USA}
\author{Nour Skaf}
\affiliation{Department of Astronomy and Astrophysics, University of California-Santa Cruz, Santa Cruz, CA, USA}
\author{Francois Bouchy}
\affiliation{Observatoire Astronomique de l’Université de Genève, Geneva, Switzerland}
\author{Isabelle Boisse}
\affiliation{Laboratoire d'Astrophysique de Marseille, Marseille, France}
\affiliation{Observatoire de Haute-Provence, 04870 Saint-Michel-l'Observatoire, France}

\author{Erica Dykes}
\affiliation{Department of Physics and Astronomy, University of Texas-San Antonio, San Antonio, TX, USA}
\author{N. Jeremy Kasdin}
\affiliation{Department of Mechanical Engineering, Princeton University, Princeton, NJ, USA}
\author{Motohide Tamura}
\affil{Astrobiology Center of NINS, 2-21-1, Osawa, Mitaka, Tokyo, 181-8588, Japan}
\affiliation{National Astronomical Observatory of Japan, 2-21-2, Osawa, Mitaka, Tokyo 181-8588, Japan}
\affiliation{Department of Astronomy, Graduate School of Science, The University of Tokyo, 7-3-1, Hongo, Bunkyo-ku, Tokyo, 113-0033, Japan}


\shortauthors{El Morsy et al.}
\begin{abstract}
We present follow-up SCExAO/CHARIS $H$ and $K$-band (R $\sim$ 70) high-contrast integral field spectroscopy and Keck/NIRC2 photometry of directly-imaged brown dwarf companion HD 33632 Ab and new radial-velocity data for the system from the SOPHIE spectrograph, complemented by Hipparcos and Gaia astrometry.  These data enable more robust spectral characterization compared to lower-resolution spectra from the discovery paper and more than double the available astrometric and radial-velocity baseline.   HD 33632 Ab's spectrum is well reproduced by a field L8.5--L9.5 dwarf.   Using the Exo-REM atmosphere models, we derive a best-fit temperature, surface gravity and radius of $T_{\rm eff}$ = 1250 $K$, log(g) = 5, and $R$ = 0.97 $R_{\rm J}$ and a solar C/O ratio.   Adding the SOPHIE radial-velocity data enables far tighter constraints on the companion's orbital properties (e.g. $i$=${47.5}_{-4.7}^{+2.5}$$^{o}$) and dynamical mass (${52.8}_{-2.4}^{+2.6}$$M_{\rm J}$) than derived from imaging data and \textit{Gaia} eDR3 astrometry data alone.  HD 33632 Ab should be a prime target for multi-band imaging and spectroscopy with the \textit{James Webb Space Telescope} and the \textit{Roman Space Telescope}'s Coronagraphic Instrument, shedding detailed light on HD 33632 Ab's clouds and chemistry and providing a key reference point for understanding young exoplanet atmospheres.
\end{abstract}

\section{Introduction}
In the past decade, large direct imaging campaigns using \textit{extreme} adaptive optics (extreme AO) systems coupled with near-infrared (IR) integral field spectrographs have discovered jovian planets and brown dwarfs on solar system scales and characterized their atmospheres \citep[e.g.][]{Macintosh2015,Chauvin2017,Konopacky2016,Rajan2017,Cheetham2019,Currie2023a}.  Most of these programs select stars for imaging observations based on properties like age and distance without any direct dynamical evidence for a companion \citep{Nielsen2019,Vigan2021,Currie2023b}.   Unfortunately, despite surveying hundreds of stars, such ``blind" surveys have resulted in a low yield of new substellar companion discoveries: e.g. two new companions out of 300 targeted with the Gemini Planet Imager Exoplanet Survey.    

\begin{deluxetable*}{lllllllll}[ht]
     \tablewidth{0pt}
    \tablecaption{HD 33632 Observing Log\label{obslog}}
    \tablehead{\colhead{UT Date} & \colhead{Instrument} &  \colhead{Seeing$^{c}$ (\arcsec{})} &{Filter} & \colhead{$\lambda$ ($\mu m$)$^{a}$} 
    & \colhead{$t_{\rm exp}$} & \colhead{$N_{\rm exp}$} & \colhead{$\Delta$PA ($^{o}$)} & SNR$^{d}$ (HD 33632 Ab)}
    \startdata
    20211018 & SCExAO/CHARIS$^{a}$ &  0.5--0.7 & $H$ & 1.47--1.79& 60.48 & 27 & 19.28  & 27.60 \\
    20211018 & SCExAO/CHARIS$^{a}$ &  0.5--0.7 & $K$ & 2.01--2.36& 60.48 & 25 & 14.86  & 29.16 \\
    20240223 & Keck/NIRC2$^{a}$ &  0.2-0.5 & $L_{\rm p}$ & 3.78 & 30 & 50 & 26.86  & 29.88 \\
    \enddata
   For CHARIS data, the $\lambda_{\rm \mu m}$ column refers to the wavelength range.  For NIRC2 imaging data, it refers to the central wavelength. 
    \label{obslog_hd33632}
    \end{deluxetable*}

Recent studies have demonstrated the advantage of a different approach to direct imaging discovery and characterization: targeting stars showing dynamical evidence for a substellar companion from precision calibrated astrometry from the \textit{Gaia} and \textit{Hipparcos} missions.   This strategy has led to the discovery of planets orbiting HIP 99770 and AF Lep, a planet or brown dwarf around HIP 39017, and numerous other brown dwarf companions on 10--30 au orbits \citep{Currie2023a, deRosa2023,Mesa2023, Tobin2024, Franson2023, Li2023,Kuzuhara2022,Swimmer2022}.  Discovery yields from these surveys are up to 5 times higher than those from unbiased surveys (El Morsy et al. 2024 submitted). 

A focus on direct imaging these \textit{accelerating} stars also enables more in-depth characterization studies. Direct imaging data by itself cannot constrain the mass of a planet or brown dwarf.  Relative astrometry for most planets and brown dwarfs imaged cover a small fraction of their orbits, resulting in characteristically poor constraints on the companions' semimajor axes, inclinations, eccentricities, and other parameters \citep[e.g. see][]{Bowler2020}. However, jointly analyzing the relative astrometry of a companion from imaging with the absolute astrometry of the star from \textit{Gaia} and \textit{Hipparcos} can directly yield dynamical masses and precise orbital parameters \citep[e.g.][]{Brandt2021b,Brandt2021c}.   

HD 33632 Ab is a $\sim$20 au-separation brown dwarf companion orbiting a Sun-like star and one of the first examples of a joint direct imaging and astrometric discovery of a substellar companion.  The companion was discovered by \citet{Currie2020a} from direct imaging and spectroscopy data obtained with SCExAO/CHARIS\footnote{SCExAO stands for the Subaru Coronagraphic Extreme Adaptive Optics Project; CHARIS refers to the Coronagraphic High Angular Resolution Imaging Spectrograph \citep{Jovanovic2015,Groff2016}.} in the near-infrared (IR), imaging from Keck/NIRC2 in the thermal IR, and astrometric data from the \textit{Hipparcos-Gaia Catalogue of Accelerations} \citep[HGCA][]{Brandt2021b}.   Its spectrum matches that of an L9.5$^{+1.0}_{-3.0}$ object: near the L/T transition probing the dissipation of clouds at high altitude in substellar atmospheres \citep[e.g.][]{Saumon2008}.   Cross-correlating very high spectral resolution data with templates identifies water and carbon monoxide lines in its atmosphere and places some constraints on the companion's carbon-to-oxygen ratio \citep{Hsu2024}.   Joint modeling of the direct imaging data, astrometric data, and archival radial-velocity (RV) constrained the mass to be $\sim$ 46 $\pm$ 8 $M_{\rm J}$.  Updated modeling using \textit{Gaia eDR3} astrometry more precisely estimates  the companion mass (50$^{+5.6}_{-5.0}$ $M_{\rm J}$) and likewise improves orbital constraints (e.g. $i$ = 45.2$^{+4.7}_{-11}$ degrees vs. 39.4$^{+8}_{-20}$ degrees) \citep{Brandt2021c}.  

With a near-IR companion-to-star contrast of only $\sim$5$\times$10$^{-5}$, an angular separation of $\approx$0\farcs{}7, and clear orbital motion detected over two years, HD 33632 Ab is particularly amenable to follow-up dynamical and atmospheric characterization.   As shown in \citet{Currie2021}, two epochs of HD 33632 Ab astrometric points from direct imaging separated by $\sim$2 years from imaging reduce the companion mass uncertainty by 50\% compared to a single epoch of imaging data.   Additional epochs may further constrain the companion's mass and orbital properties.  Because HD 33632 Ab is an old, relatively inactive F8V star \citep{Currie2020a}, it is well suited for additional precision RV measurements,  complementing astrometric data.   The CHARIS data in the discovery paper consist of integral field spectroscopy (IFS) at a very low resolution ($R$ $\sim$ 20).  Higher-resolution IFS detections of HD 33632 Ab may enable a more precise characterization of the companion's atmospheric properties like gravity and carbon chemistry \citep[e.g.][]{Barman2011}.

In this study, we present the first comprehensive atmospheric and dynamical follow-up characterization of HD 33632 Ab, combining new direct imaging and RV data with HGCA astrometry to constrain the companion's atmosphere, orbit, and mass.
Direct imaging data from SCExAO/CHARIS at $H$ and $K$ band at higher resolution bet: these data and Keck/NIRC2 $L_{\rm p}$ imaging nearly triple HD 33632 Ab's astrometric baseline.  Similarly, new data from the SOPHIE spectrograph double the RV time baseline compared to data presented in \citet{Currie2020a}.  
\begin{figure*}[ht!]
    \includegraphics[width=0.33\textwidth]{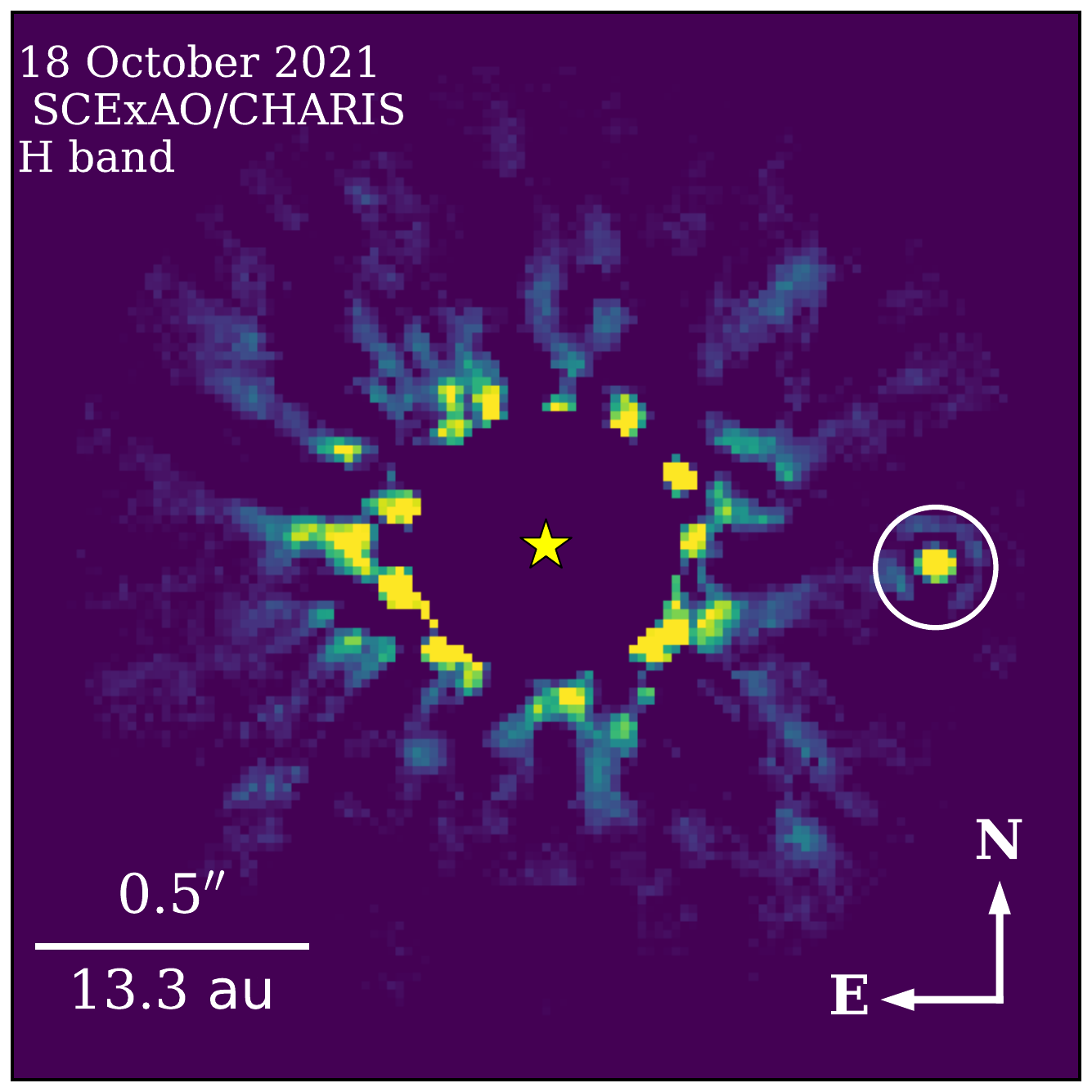}%
    \includegraphics[width=0.33\textwidth]{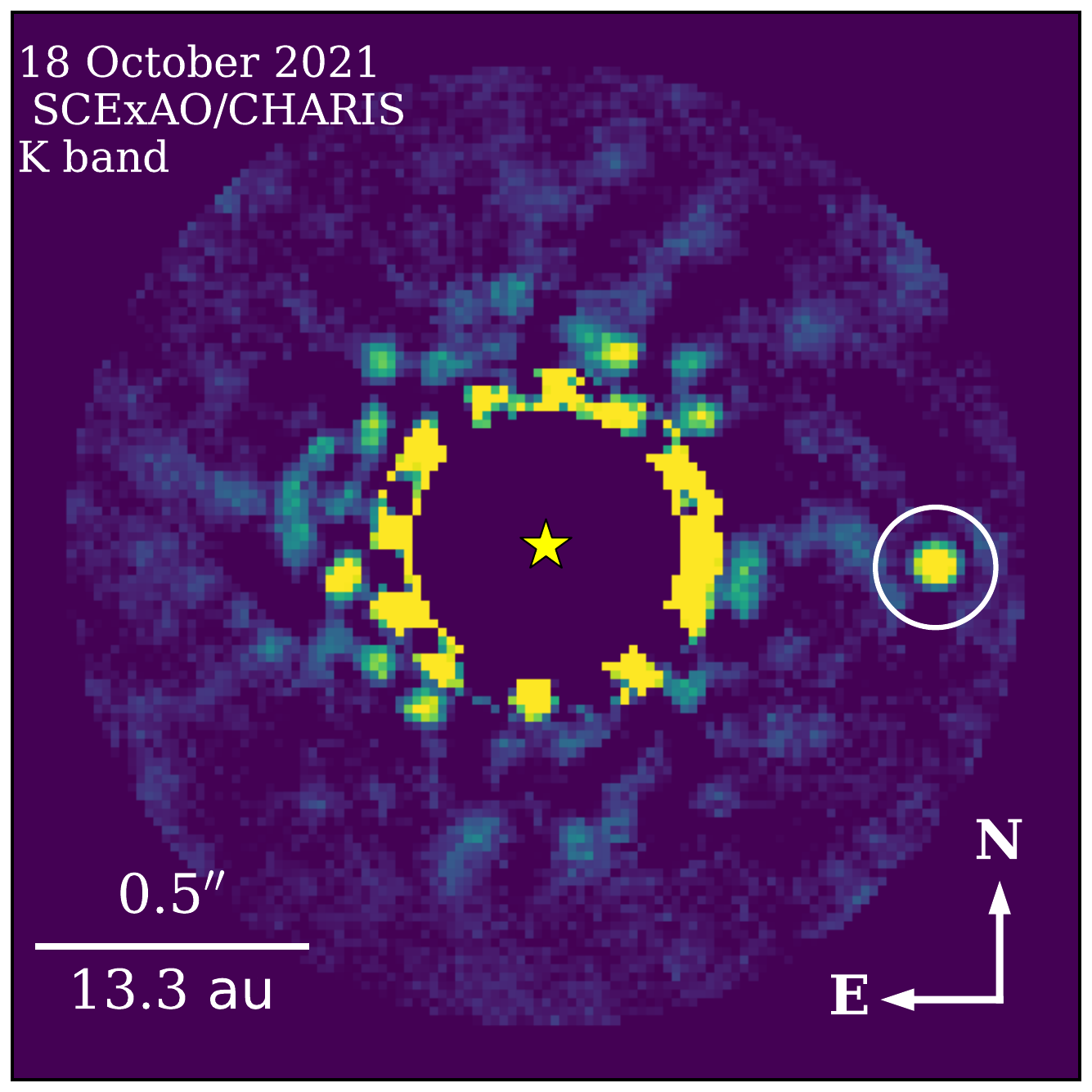}%
    \centering
    \includegraphics[width=0.33\textwidth]{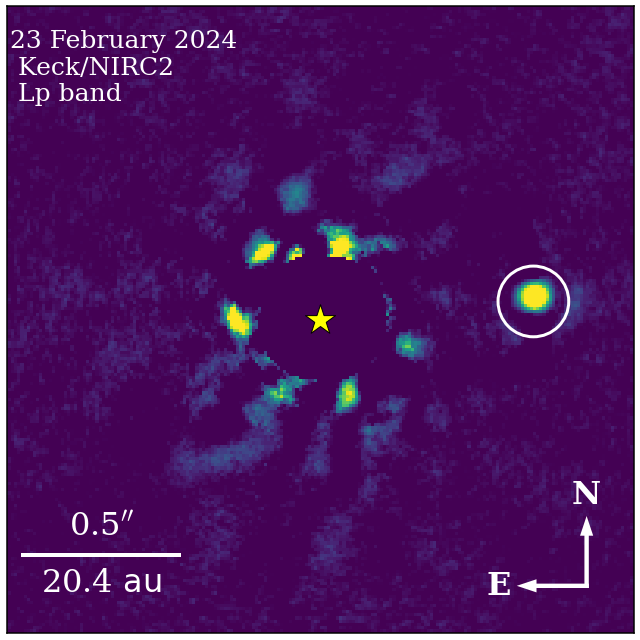}
  \caption{
  Detection of HD 33632 Ab in $H$ and $K$ band from 18 October 2021 SCExAO/CHARIS data (left, middle) and Keck/NIRC2 in $L_{\rm p}$ from 23 February 2024 (right).  The companion (circled in white) is located at $\rho \approx 0.73 \arcsec{}$ from the star in the CHARIS data and $\rho \approx 0.68 \arcsec{}$ in the Keck/NIRC2 data. 
  }
  \label{fig:images}
\end{figure*}

\section{Data}
\subsection{ High-Contrast Imaging Observations and Data Reduction}

\subsubsection{SCExAO/CHARIS Data}
We obtained follow-up observations of HD 33632 A with SCExAO/CHARIS in the $H$ and $K$ band ``high resolution" modes (R $\sim$ 60--70) on 18 October 2021 (Table \ref{obslog_hd33632}).   Conditions were photometric and with moderate windspeeds ($\sim$ 15 mph), resulting in average seeing ($\theta_{\rm V}$ $\sim$ 0\farcs{}5--0\farcs{}7) for Maunakea.   We targeted HD 33632 for a total of $\approx$ 1 hour clock time ranging between $\sim$ 1.2 and 2.3 hours after transit, alternating between the $H$ and $K$ filters. 

All CHARIS data consisted of 60.48 $s$ exposures obtained in \textit{angular differential imaging}(ADI)/pupil tracking mode, allowing the sky to rotate on the detector with time \citep{Marois2006}.  The total integration time and parallactic angle motion for $H$ and $K$ were 27.2 minutes, 19.28$^{o}$ and 25.2 minutes, 14.87$^{o}$, respectively.  To provide astrometric and spectrophotometric calibration, we modulated the SCExAO deformable mirror with an amplitude of 25 nm to generate satellite spots \citep[][]{Jovanovic2015-astrogrids}.  For all data, we inserted a Lyot coronagraph into the focal plane with a 0\farcs{}139 occulting spot diameter, suppressing the stellar halo.

To extract data cubes from the raw CHARIS frames, we used the pipeline from \citet{Brandt2017}.  
Subsequent basic processing followed standard steps utilized for reducing low-resolution/broadband data using the CHARIS Data Processing Pipeline \citep{Currie2020b}.   For the K-band data,  we used blank sky exposures to subtract thermal background emission.   For both passbands, we then registered each cube to a common center and spectrophotometrically calibrated the data assuming a Kurucz atmosphere model appropropriate for an F8V star equal to HD 33632 A's brightness ($m_{\rm H, K_{\rm {s}}}$ = 5.19 $\pm$ 0.02, 5.17 $\pm$ 0.02).  

Inspection of the extracted CHARIS cubes shows that, like the low-resolution data presented in \citet{Currie2020a}, HD 33632 Ab is faintly visible in many individual channels even without point-spread function (PSF) subtraction, especially in $K$ band.   Combined with the modest parallactic angle rotation enabled by our observations, we therefore tuned our point-spread function (PSF) subtraction approach towards conservative settings resulting in minimal spectrophotometric and astrometric biasing, not to maximize signal-to-noise.  We used the Adaptive, Locally-Optimized Combination of Images (A-LOCI) approach \citep{Currie2012,Currie2015} employing a pixel mask over the subtraction zone \citep{Marois2010b,Currie2012} and constructing a reference PSF by minimizing the residuals within the optimization zone \citep[see][]{Lafreniere2007}.  
This approach results in less signal loss and -- when compared with forward-modeling without masking e.g. as with KLIP -- significantly reduces the possibility that the companion's bright signal complicates forward-modeling techniques needed to extract accurate spectrophotometry \citep{Currie2018}. 
Our reductions used a \textit{singular value decomposition} (SVD) cutoff of 10$^{-6}$ and a rotational gap of $\delta$ = 0.75 PSF footprints.

\subsubsection{Keck/NIRC2 Data}
We also obtained follow-up Keck II imaging of HD 33632 with the
NIRC2 data in the $L_{\rm p}$ broadband filter ($\lambda_{o}$ = 3.78 $\mu m$) using Keck's facility AO system.   Conditions were exceptional for thermal IR AO imaging, with seeing reaching as low as 0\farcs{}20, minimal winds, and less than 5\% humidity.   We used a Lyot coronagraph with a 0\farcs{}2 diameter pupil mask and also observed in ADI mode for a total of 25 minutes of integration time and $\sim$27$^{o}$ of parallactic angle motion.  For photometric calibration, we observed the bright star HD 32537 just prior to our HD 33632 sequences.

We reduced the NIRC2 data using the ADI-based pipeline from \citet{Currie2011}.  Key steps included sky subtraction, registering each image to a common center, photometric calibration, spatial filtering, PSF subtraction (with A-LOCI), and forward-modeling.   Our algorithm settings are marginally more aggressive -- identical rotation gap but no pixel masking -- as HD 33632 Ab was less clearly visible in raw data than with CHARIS.

\begin{figure}
\centering
   \includegraphics[width=0.5\textwidth]{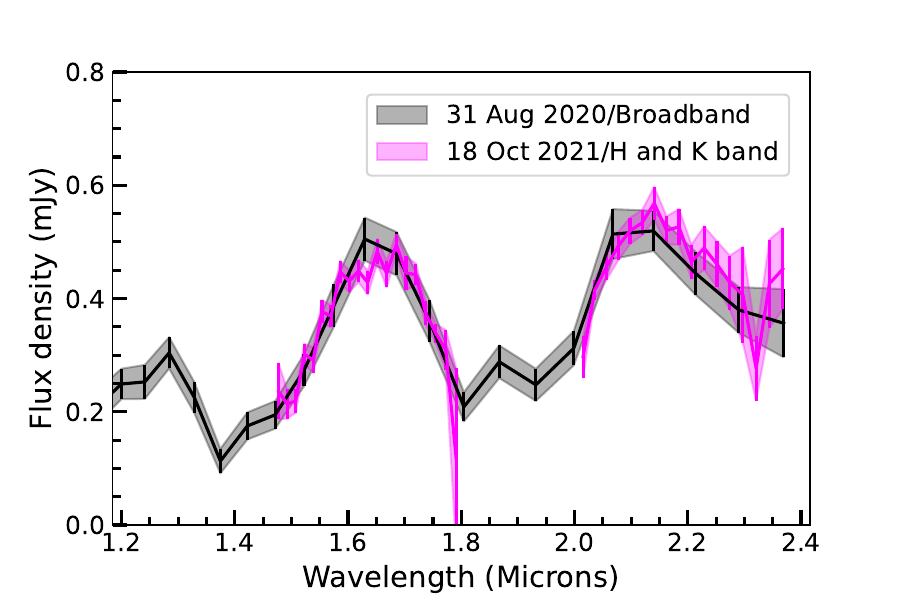}
   \caption{Our October 2021 CHARIS $H$ and $K$ band spectra of HD 33632 Ab (magenta) compared to lower-resolution CHARIS spectra from \citet{Currie2020a} taken in 2020 (black).  Vertical bars denote 1-$\sigma$ errors. }
   \label{fig:spectra} 
\end{figure}

\begin{deluxetable}{llll}
 \label{astrom}
     \tablewidth{0pt}
     \tabletypesize{\scriptsize}
      \setlength{\tabcolsep}{1pt}
     \tablecaption{HD 33632Ab Astrometry}
     \tablehead{\colhead{UT Date} & \colhead{Instrument}  & \colhead{Filter} & \colhead{[E,N]$\arcsec{}$}}
     \startdata
     \textbf{Previous Data}\\
    20181018  & SCExAO/CHARIS & $JHK$ & ${-0.761, -0.176} \pm 0.005, 0.004$\\
    20181101 & Keck/NIRC2 & $L_{\rm p}$ &${-0.753, -0.178} \pm 0.005, 0.005$\\
    20200831-20200901 & SCExAO/CHARIS & $JHK$ & ${-0.740, -0.095} \pm 0.005, 0.003$\\
        \textbf{New Data}\\
    20211018 & SCExAO/CHARIS & $HK$ & ${-0.728, -0.034} \pm 0.005, 0.005$\\
    20240223 & Keck/NIRC2 & $L_{\rm p}$ & ${-0.678,+0.068} \pm 0.003, 0.003$\\
    \enddata
\end{deluxetable}
\subsubsection{Detections}
Figure \ref{fig:images} shows the detection of HD 33632 Ab in $H$, $K$, and $L_{\rm p}$ bands located at 3 o'clock from the star.  
The estimated signal-to-noise ratios (SNR) for the collapsed cubes in CHARIS $H$ and $K$ band and NIRC2 $L_{\rm p}$ filter are 27.6, 29.16, and 29.88 respectively: slightly smaller than the highest-quality broadband CHARIS dataset from \citet{Currie2020a}.   The first and last channels for each CHARIS data set are contaminated by telluric emission and thus generally have low throughput and low SNR.   Aside from these, typical detection significances are SNR $\sim$ 10-25 per channel.



\subsection{High-Contrast Imaging Spectral Extraction and Astrometry}

To correct for astrometric and spectrophotometric biasing of HD 33632 Ab due to processing, we employed forward-modeling as described in \citet{Currie2018}, treating the companion signal as a small perturbation on the LOCI coefficients used to construct the reference PSF \citep[see also][]{Pueyo2016}.   Forward-modeling reveals low signal loss ($\sim$12-20\% per channel) and negligible astrometric biasing.   Figure \ref{fig:spectra} compares the CHARIS $H$ and $K$ band spectra with the broadband, lower-resolution spectrum from \citet{Currie2020a}; 
Appendix A lists the spectra's the flux density, uncertainty in the flux density, and SNR in each channel.  The spectra agree to within errors at all channels except at $\sim$1.8 $\mu m$ region affected by telluric absorption.   



\begin{figure}[ht!]
   
    \includegraphics[width=0.5\textwidth,clip]{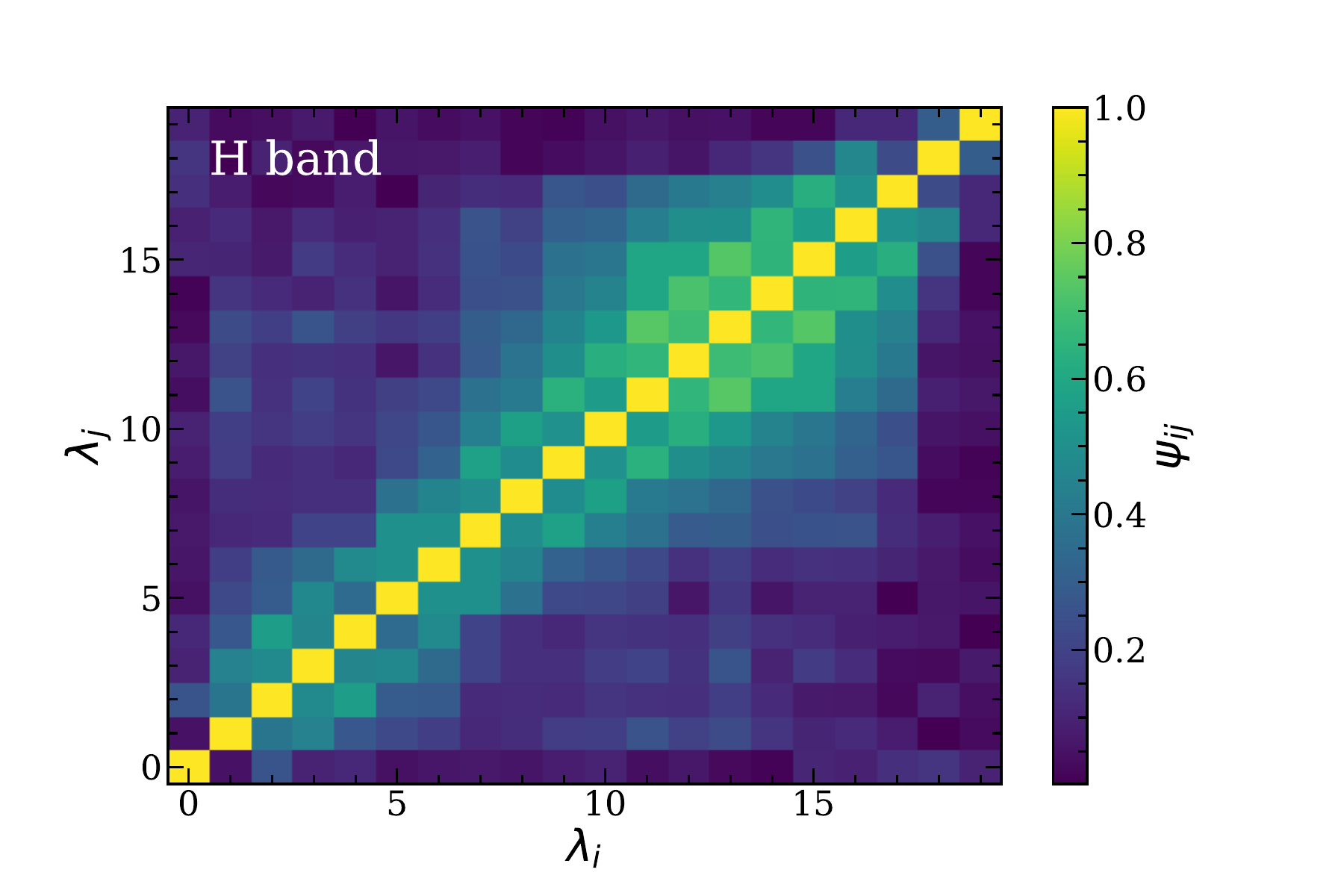}\\
    \includegraphics[width=0.5\textwidth,clip]{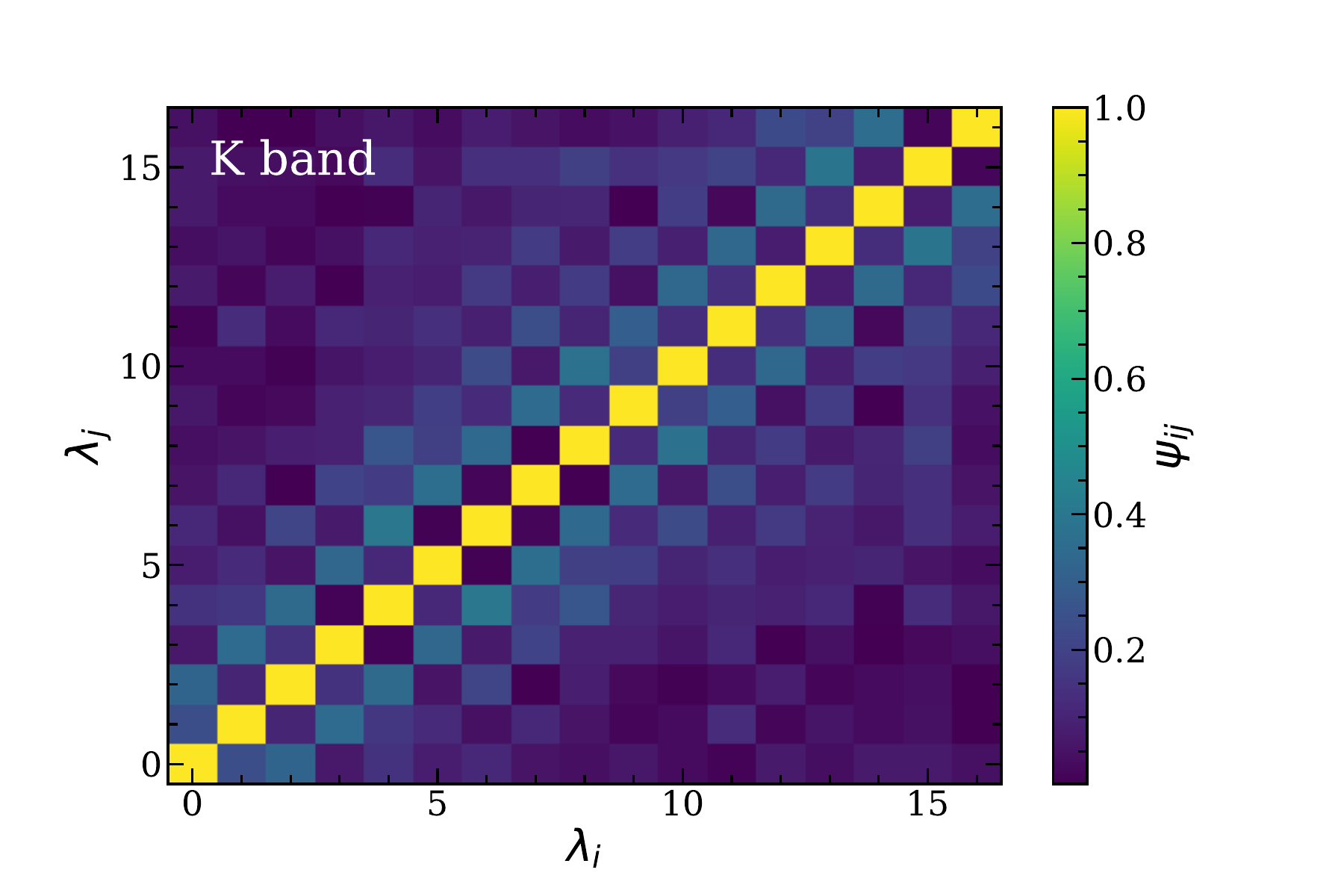}%
  \caption{Correlation matrices in H (upper panel) and K (lower panel) bands as a function of spectral channel. The off-diagonal elements identify spectrally correlated noise.}
  \label{fig:corr_matrix}
\end{figure}

Figure \ref{fig:corr_matrix} displays the spectral covariance matrices \citep{GrecoBrandt2016} for H (top panel) and K band (bottom panel).  The off-diagonal elements represent the spectral and spatial correlation of residuals, identifying spectrally correlated noise.  For $H$ band, residuals in a given channel are typically correlated with those $\pm$ 2 channels away at a level of $\psi$ $\gtrsim$ 0.5.   Covariances are weaker in K band.


We derive $H$ and $K_{\rm s}$ band photometry by convolving the CHARIS $H$ and $K$-band spectra with the Mauna Kea Observatories filter transmission \citep[e.g.][]{Currie2018}.   HD 33632 Ab's $H$ and $K_{\rm s}$ band apparent magnitudes are 16.08 $\pm$ 0.06 and 15.36 $\pm$ 0.06, respectively, and its $H$-$K_{\rm s}$ color is then 0.72 $\pm$ 0.08.   For NIRC2 $L_{\rm p}$ photometry, we used the bright star HD 32537 for photometric calibration, yielding m$_{\rm L_{\rm p}}$ = 13.64 $\pm$ 0.06.  These values agree with ones derived from lower-resolution broadband spectra from \citet{Currie2020a} to within 1-$\sigma$ for each measurement.

Table \ref{astrom} lists our astrometry and previous measurements from \citet{Currie2020a}.  We adopt the astrometric calibration from \citet{Currie2022} appropriate for CHARIS observations obtained through 2021.  HD 33632 Ab's position with CHARIS obtained from averaging the $H$ and $K$ band astrometry is estimated to be at [E,N]$\arcsec{}$ = [$-$0\farcs{}728,$-$0\farcs{}034] $\pm$ [0\farcs{}005,0\farcs{}005]. The astrometric errors consider algorithm biasing (negligible), the intrinsic SNR of the detection, centroiding uncertainty (set to 0.25 pixels or $\sim$0\farcs{}004), and differences in the $H$ and $K$ band measurements.   The NIRC2 astrometry was measured to be [E,N]$\arcsec{}$ = [$-$0\farcs{}678,0\farcs{}068] $\pm$ [0\farcs{}005,0\farcs{}005].   Compared to its 2020 position, HD 33632 Ab has shifted closer to its host star by a displacement of $\Delta$([E,N]) $\sim$ [$-$0\farcs{012},0\farcs{}061] in the CHARIS data and $\Delta$([E,N]) $\sim$ [$-$0\farcs{062},0\farcs{}163] in the NIRC2 data.

\subsection{Radial-Velocity Data From the SOPHIE Archive}

To our high-contrast imaging data, we add precision radial-velocity (RV) data from the SOPHIE instrument \citep{Bouchy2006,Perruchot_2008_SOPHIE} and its upgraded version SOPHIE+ \citep{Bouchy2013}. The data were taken between 2006 October 14 and 2018 December 6 and extend the RV coverage (1998-1-18 to 2009-2-1) from \citet{Currie2020a}, which was drawn entirely from the Lick Observatory archive data using the Hamilton spectrograph \citep{Fischer2014}.  The spectra were taken in high-resolution mode using the Thorium-Argon lamp for wavelength determination.
The RV drift caused by a spectrograph instability was monitored by simultaneously observing Thorium-Argon lamp or Fabry-Pérot étalon.
We assumed that an offset between the RVs taken by SOPHIE and SOPHIE+ is insignificant based on the previous studies performed with those instruments \citep[e.g.,][]{Bouchy2013, Heidari_2024_SOPHIE,Diaz_2016_SOPHIE,Kiefer_2019_SOPHIE}.
Exposure times ranged between 120 $s$ and 1200 $s$.  

We downloaded the wavelength-calibrated 1D spectra of HD 33632A from the SOPHIE archive\footnote{\url{http://atlas.obs-hp.fr/sophie/}}. 
To measure RVs of HD 33632\,A the spectrum data were analyzed using the SpEctrum Radial Velocity AnaLyser (SERVAL) pipeline \citep{Zechmeister2018}, which can yield a typical RV precision of 1 $m s^{-1}$. 
When running SERVAL, we ignored spectrum orders in high dispersion spectrscopy if their SNRs are smaller than 10 and set the absolute velocity of HD 33632A to be $-$1.705 k m$^{-1}$ according to \citet{Soubiran_2018}.

\par 
Measurements of relative RVs still suffer from imperfect calibrations, varying nightly zero point (NZP) in the RV measurements.
The NZP variation of SOPHIE has been monitored by measuring RVs of servarl RV-stable stars in each observing night\citep{Courcol_2015_SOPHIE,Hara_2020_SOPHIE}.
Recently, \citet{Grouffal_2024_SOPHIE_NZP} developed a new method to model the NZP drifts by applying Gaussian process (GP) regressions to the RV measurements of the stable stars.
The NZP corrections from the GP modeling and the observing epoch corresponding to the NZPs were tablulated in \citet{Grouffal_2024_SOPHIE_NZP}, which we adopt in this study. 
If an observation of HD 33632\,A was conducted within one day from an epoch in the NZP table, we adopted the NZP correction of that epoch.
Also, the time-nearest NZP correction was adopted if multiple NZP corrections were listed from this selection.
We propagated the uncertainties of NZP corrections, which are also tabulated in \citet{Grouffal_2024_SOPHIE_NZP}, to the errors of NZP-corrected RVs of HD 33632\,A. 
In contrast, if this selection provided no available NZP correction for an observing epoch, we did not correct for the NZP in that epoch but instead inflated the RV errors by adding the scatter ($\simeq$1.7 m s$^{-1}$) of all the NZP values in quadrature.  
Appendix B lists our extracted RVs and uncertainties.   The spectra have a mean RV uncertainty of $\sim$1.8 $m s^{-1}$.
Between October 2014 (JD = 2456939.5) and December 2018 (JD = 2458459.5), the SOPHIE data suggests a long-term RV trend, consistent with a possible RV detection of HD 33632 Ab.

\section{Atmospheric Analysis}

\subsection{Empirical Comparisons}


We compared HD 33632 Ab's H and K band spectra to those from MLT dwarfs in the Montreal Spectral Library \citep{Gagne2015, Robert2016, Delorme2012, Naud2014}. We compare each library spectrum to HD 33632 Ab's by computing the $\chi^{2}$ value as the sum of values in $H$ and $K$ band, separately: 

\begin{equation}
    \begin{multlined}
    \chi^2 = \sum_i ( f_{\nu,i} - \alpha F_{\nu,i(H)} )^T C_i(H)^{-1} ( f_{\nu,i(H)} - \alpha F_{\nu,i(H)} )\\
    + \sum_j ( f_{\nu,j} - \alpha F_{\nu,j(K)} )^T C_j(K)^{-1} ( f_{\nu,j(K)} - \alpha F_{\nu,j(K)} ),
     \end{multlined}
    \label{eq:chi2calc}
\end{equation}

where we choose the scaling coefficient $\alpha$ that minimizes $\chi^{2}$.  $F_{\nu}$ and $f_{\nu}$ are the flux of the template and target spectra, respectively, in the $i^{th}$ H band wavelength bin or $j^{th}$ K band wavelength bin.  The H and K band spectral covariances are $C_(H)$ and $C_(K)$.  

\begin{figure}
\centering
   \includegraphics[width=0.5\textwidth,trim=12mm 5mm 2mm 10mm,clip]{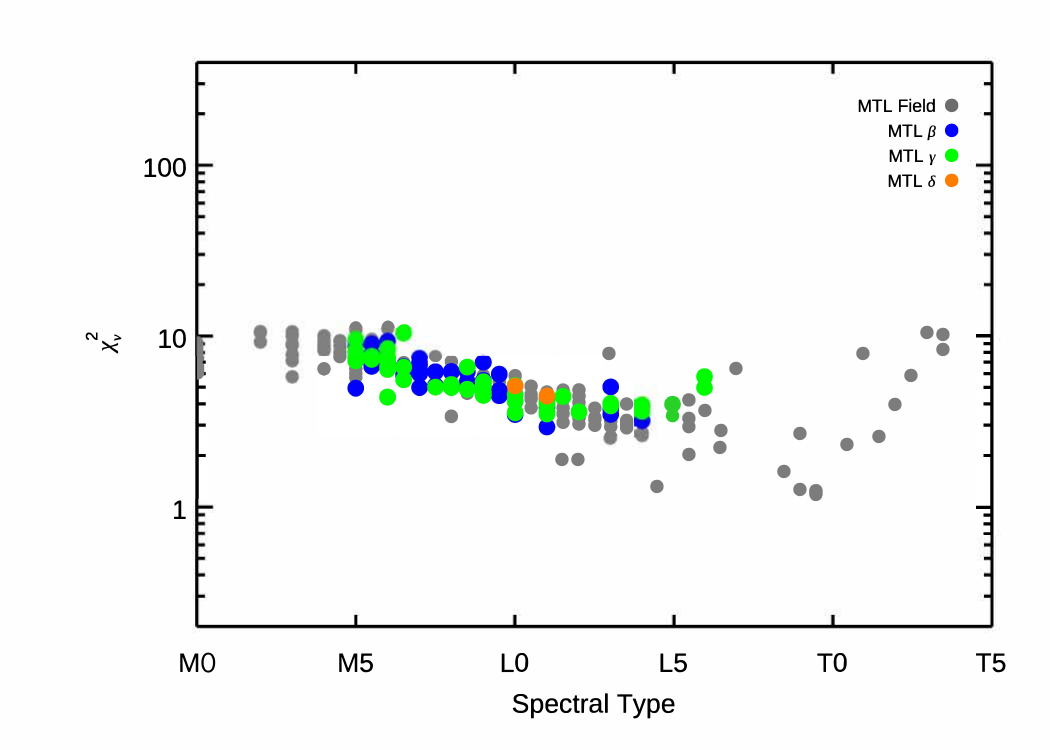}
   \caption{The $\chi_{\nu}^{2}$ distribution comparing the spectrum of HD33632 Ab to spectra of objects included in the Montreal Spectral Library with varying gravity levels (field in grey, intermediate in blue, low in green, and very low in orange).}
   \label{fig:chi_square} 
\end{figure}

\begin{figure}[ht!]
      \includegraphics[width=0.5\textwidth,clip]{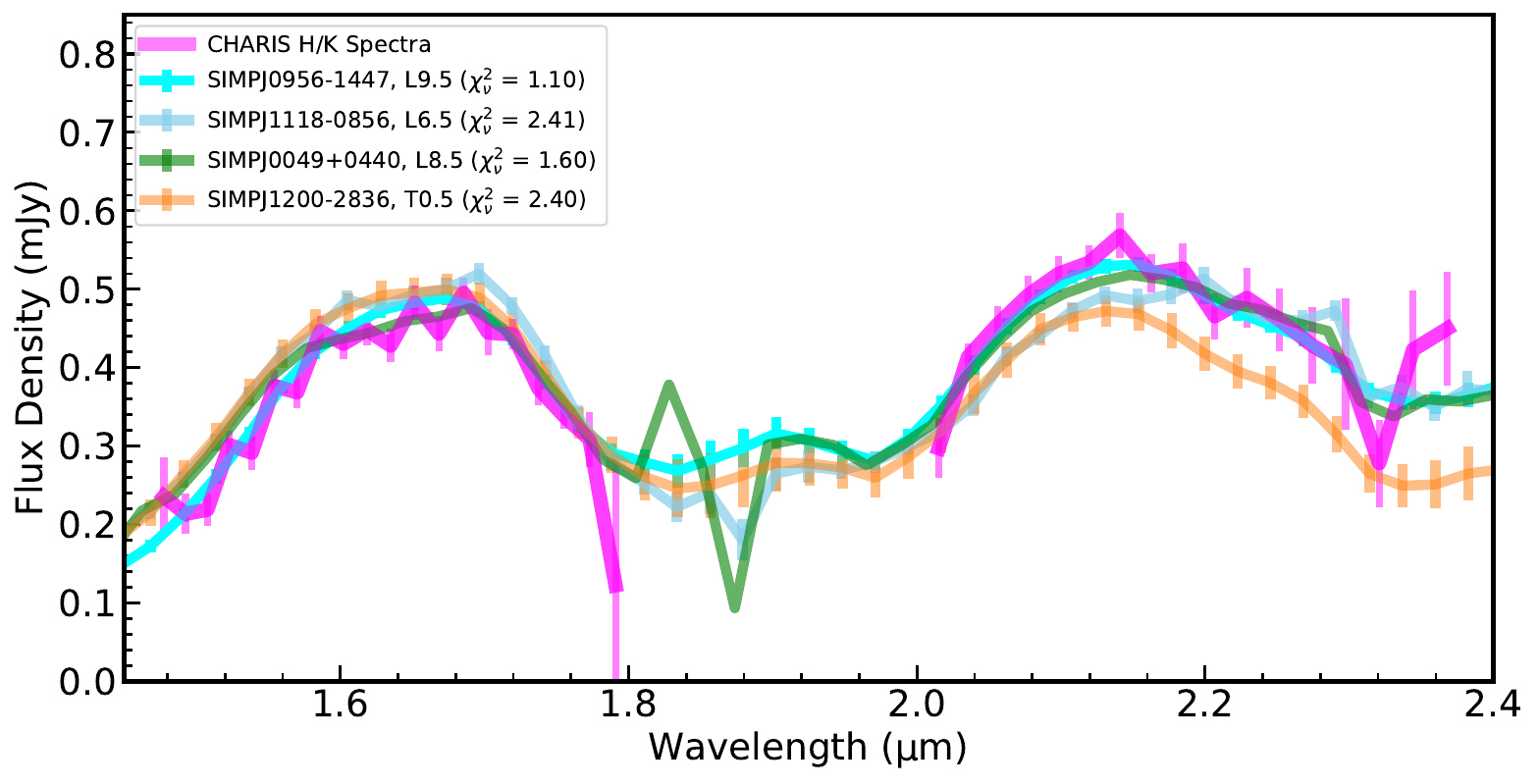}\\
    \includegraphics[width=0.5\textwidth,clip]{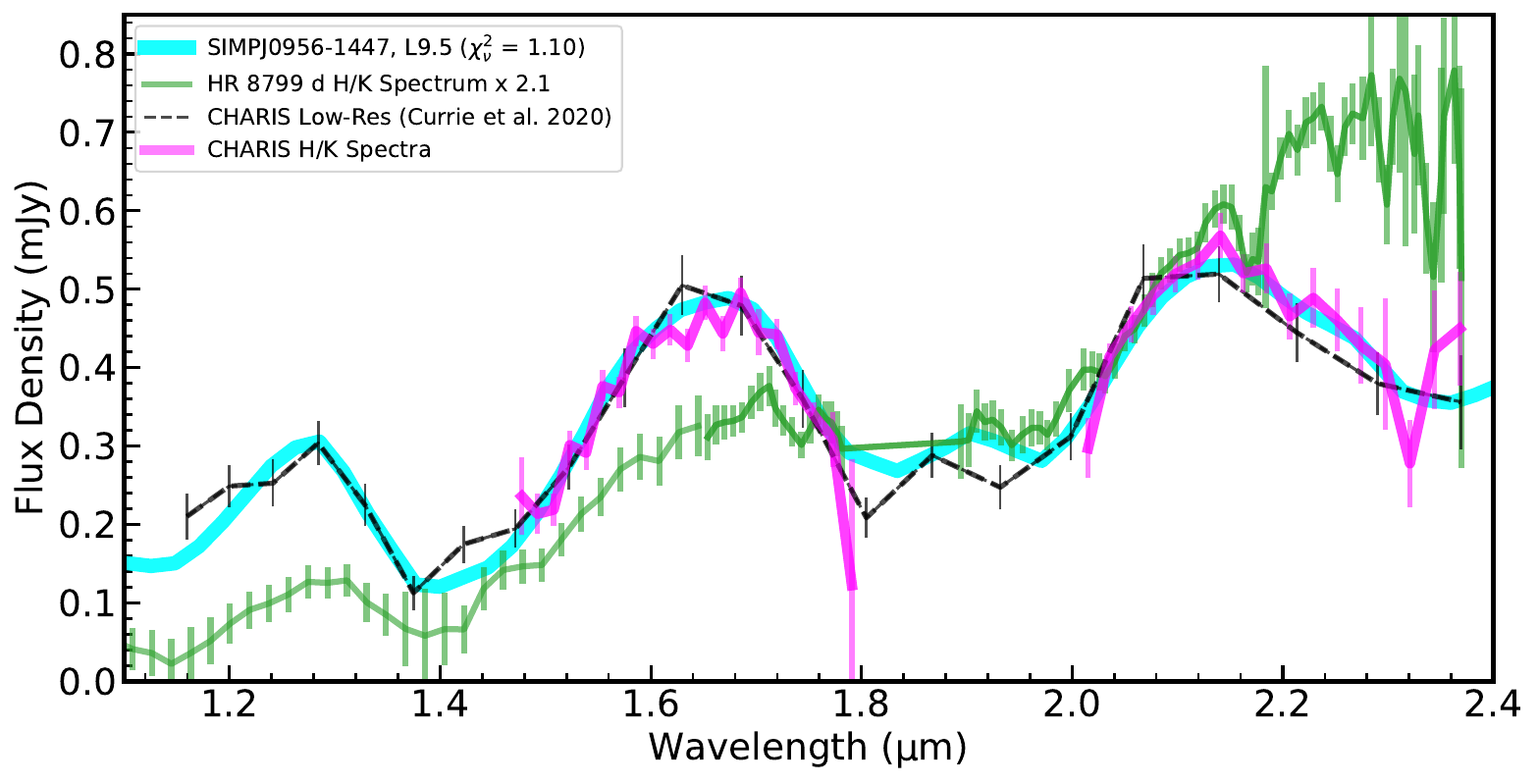}
  \caption{Comparison between the HD 33632 Ab H and K band spectra and other objects. (top panel) Comparisons between HD 33632 Ab's $H$ and $K$-band spectra, SIMPJ0956-1447 (the best-fit brown dwarf in the Montreal Spectral Library), and best-fitting L8.5 and T0.5 dwarfs.  (bottom panel) Comparisons with the CHARIS low-resolution spectrum from \citet{Currie2020a}, SIMPJ0956-1447, and the HR 8799 d spectra drawn from \citet{Zurlo2016} and \citet{Greenbaum2018} over 1.1--2.4 $\mu m$.}
  \label{fig:empcomp}
\end{figure}

Fig \ref{fig:chi_square} shows that our comparisons reach a $\chi^{2}$ minimum for the L9.5 spectral type, where the best-fitting object is the field L9.5 dwarf SIMPJ0956-1447 \citep{Robert2016}, which was also the best fit to the CHARIS low-resolution spectrum from \citet{Currie2020a}.   The top panel of Figure \ref{fig:empcomp} demonstrates how earlier and later spectral types provide a poorer match to HD 33632 Ab.  Consistent with results from \citet{Currie2020a} for low-resolution $JHK$ spectra, HD 33632 Ab's $H$ and $K$ band spectra best represent an L9.5$^{+1.0}_{-3.0}$ dwarf, where spectral types between L8.5 and L9.5 are favored\footnote{Figure \ref{fig:chi_square} shows one L4.5 dwarf also with a low $\chi_{\nu}^{2}$ value of $\approx$ 1.3: SIMPJ1122+0343.  However, this object's spectrum's spectrum is particularly noisy compared to most others in the Montreal Spectral Library.  Other L4--L5 dwarfs provide a poor match to HD 33632 Ab's spectrum.}.

The bottom panel of Figure \ref{fig:empcomp} compares the HD 33632 Ab $H$ and $K$ band spectra to this object, the low-resolution HD 33632 Ab CHARIS spectrum from \citet{Currie2020a}, and $H$ and $K$ band spectra for HR 8799 d from \citet{Zurlo2016} and \citet{Greenbaum2018}  scaled to HD 33632 Ab.   SIMPJ0956-1447 is an excellent match to the HD 33632 Ab $H$ and $K$ band spectra ($\chi_{\nu}^{2}$ $\sim$ 1.1) and likewise the $J$ band portion of the low-resolution spectrum.  In contrast, the HR 8799 d spectrum is flatter and redder, fainter at $J$ and $H$ band but significantly brighter at $\lambda$ $>$ 2.2 $\mu m$ sensitive to methane absorption.   



\subsection{Atmospheric Modeling}

\begin{deluxetable*}{llllll}[ht]
     \tablewidth{0pt}
    \tablecaption{Atmosphere Model Parameter Space\label{atmosgrid}}
    \tablehead{\colhead{Model Grid} & \colhead{$T_{\rm eff}$} &  \colhead{$log(g)$} & \colhead{Metallicity} & \colhead{carbon chemistry} & \colhead{Best Fit Parameters ($\chi_{\rm \nu}^{2}$)}} 
    \startdata
    \textit{Full Grids}\\
    BT-Settl/AGS2009 & 400-2500 & 3.5-5.5 & solar & equilibrium chemistry & $T_{\rm eff}$ = 1300 $K$, log(g) = 4.0, $R$ = 0.84 $R_{\rm Jup}$ (2.973)\\
    Exo-REM/S=0.003 & 400-2000 & 3.0-5.0 & [M/H] = -1 to 1 & non-equilbrium/ & $T_{\rm eff}$ = 1250 $K$, log(g) = 5.0, $R$ = 0.97 $R_{\rm Jup}$\\
     & & & &[C/O] = 0.1--0.8 &[M/H] = 0, [C/O] = 0.55 (1.575)\\
    \hline
    \textit{Selected Models}\\
    Lacy/Burrows/AE100 & 1000-1500 & 3.5-5.0 & solar & equilbrium chemistry & $T_{\rm eff}$ = 1400 $K$, log(g) = 4.5, $R$ = 0.74 $R_{\rm Jup}$ (2.511)\\
    Lacy/Burrows/AEE100 & 1000-1500 & 3.5-5.0 & solar & equilbrium chemistry & $T_{\rm eff}$ = 1500 $K$, log(g) = 4.5, $R$ = 0.64 $R_{\rm Jup}$ (2.386)\\
    Lacy/Burrows/AEE100 & 1000-1500 & 3.5-5.0 & solar & non-equilibrium/0.1$\times$CH$_{\rm 4}$ & $T_{\rm eff}$ = 1300 $K$, log(g) = 4.5, $R$ = 0.8 $R_{\rm Jup}$ (2.198)\\
    Lacy/Burrows/E60 & 1300-1600 & 4.0-5.0 & solar & non-equilibrium/0.1$\times$CH$_{\rm 4}$ & $T_{\rm eff}$ = 1400 $K$, log(g) = 4.5, $R$ = 0.70 $R_{\rm Jup}$ (2.634)\\
    \enddata
    \end{deluxetable*}

To further characterize HD 33632 Ab's atmosphere, we fit our CHARIS spectra and NIRC2 photometry to atmosphere models varying temperature, gravity, metallicity, and chemistry.   Large grids of atmosphere models identify the parameter space best fitting HD 33632 Ab data; selected atmosphere models from other sources explore the robustness of these conclusions to different modeling formalisms.   Table \ref{atmosgrid} summarizes the parameter space covered by each of these model sources, the parameter space they cover, and properties of the best-fit models from each source.

\subsubsection{Modeling Approach}
\begin{itemize}
\item The BT-Settl and Exo-REM Model Grids

First, we consider two grids of models: the widely-used BT-Settl models\footnote{These models were downloaded from the Theoretical Spectra Web Server: \url{http://svo2.cab.inta-csic.es/theory/newov2/ .}} and the recently-available Exo-REM models \citep{Allard2012,Charnay2018}.   The BT-Settl atmosphere models adopt abundances from \citet{Asplund2009} abundances.   Abundances sources for the Exo-REM models include the ExoMol database \citep{TennysonYurchenko2012,Yurchenko2014} and others described in \citet{Baudino2015,Baudino2017} and \citet{Charnay2018}; the Exo-REM models include non-equilibrium carbon chemistry from \citet{Charnay2018}, while the BT-Settl models assume equilibrium chemistry.

Both model grids include a self-consistent formalism for clouds and dust entrained in clouds.  The Exo-REM grid corresponds to a case with simple microphysics (iron and silicate clouds) with supersaturation parameter S=0.003, which nearly matches the near-IR color-magnitude diagram positions of field L/T transition at high gravity (log(g) = 5) and matches the reddened/underluminous L/T sequence of young, planet-mass companions at low gravity (log(g) = 3.5-4.5) \citep[see Figure 16 in ][]{Charnay2018}.

\item Lacy/Burrows Cloudy Models

Second, we consider selected models from B. Lacy and A. Burrows used in \citet{LacyBurrows2020}, as updated in in \citet{Currie2023a} to fit HIP 99770 b data.  These models cover a more limited range in temperature and gravity than the BT-Settl or Exo-REM grids.  But in contrast to these grids, the Lacy/Burrows models parameterize and vary the cloud shape function (i.e. the cloud thicknesses at a given temperature and pressure) and carbon chemistry.  Sources for the models' molecular abundances and pressure-dependent opacities are listed in the Supplementary Material from \citet{Currie2023a}

\end{itemize}

For our fits, we trim the first and last CHARIS channels for each spectrum to remove those affected by telluric absorption.   
Following \citet{Currie2023a}, we use the $\chi^{2}$ statistic to assess the fit quality to the $jth$ model:

\begin{equation}
    \begin{multlined}
    \chi_{j}^{2} = R_{H, j}^{T}C_{H}^{-1}R_{H,j} + R_{K, j}^{T}C_{K}^{-1}R_{K,j}\\
    + \sum_{i}(f_{phot,i}-\alpha_{j}~F_{phot,ij})^{2}/\sigma_{phot,i}^{2}
     \end{multlined}
    \label{eq:chi2calc2}
\end{equation}
    \begin{figure}
\centering
   \includegraphics[width=0.5\textwidth,trim=1mm 1mm 1mm 1mm,clip]{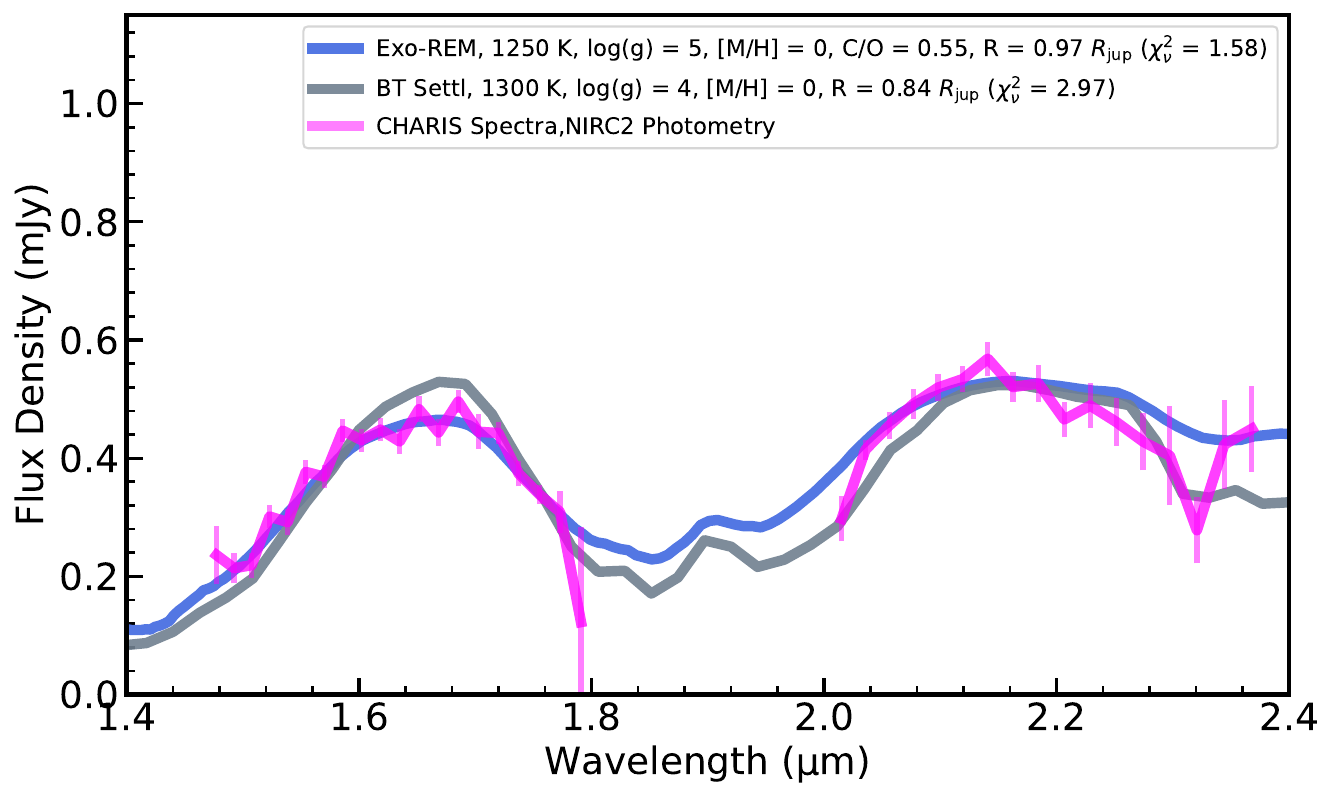}\\
   \includegraphics[width=0.5\textwidth,trim=1mm 1mm 1mm 1mm,clip]{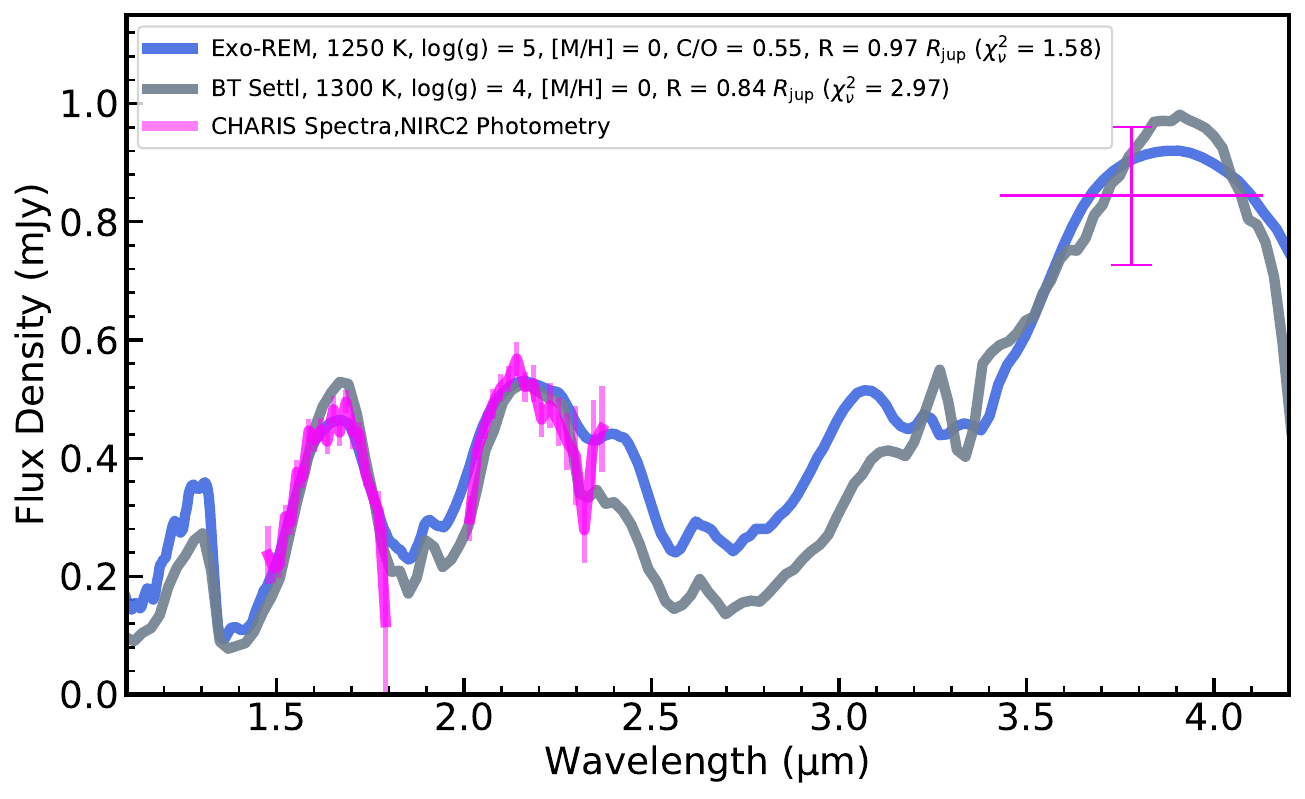}
   \vspace{-0.25in}
   \caption{HD 33632 Ab data compared to the best-fit models from the Exo-REM and BT-Settl grids, focused on the near-IR spectrum (top) and including the NIRC2 photometric point (bottom).   The horizontal bar for the NIRC2 data point depicts the bandwidth of the $L_{\rm p}$ filter.}
   \label{fig:bestfit}
\end{figure}

where the vectors $R_{H,j}$ and $R_{K,j}$ are the difference between measured ($f$) and predicted ($F$) CHARIS H and K band spectral points ($f_{spec}-\alpha_{j}F_{spec}$) and $C_{H}$ and $C_{K}$ are their covariances.  The vectors $f_{phot,i}$, $F_{phot,ij}$, and $\sigma_{phot,i}$ are measured photometry, model predicted photometry, and photometric uncertainty for the $L_{\rm p}$ measurement.The scaling factor $\alpha_{j}$ is a free parameter we vary to minimize $\chi^{2}$ and is equal to the ratio of the object radius to its distance squared ( (R/D)$^{2}$ ).  We assume a distance of 26.39 $pc$ \citep{GAIA2023}.


\begin{deluxetable}{lllllllll}[ht]
     \tablewidth{0pt}
    \tablecaption{Best-Fitting Exo-REM Models\label{exorembestfit}}
    \tablehead{\colhead{$T_{\rm eff}$ ($K$)} &  \colhead{$log(g)$} & \colhead{Metallicity} & \colhead{C/O} & \colhead{$\chi_{\rm \nu}^{2}$)}} 
    \startdata
    1250 & 5.0 & 1.0 & 0.55 & 1.575\\
    1050 & 4.5 & 1.0 & 0.55 & 1.727\\
    1100 & 5.0 & 1.0 & 0.30 & 1.784
    \enddata
    \vspace{-0.5in}
    \end{deluxetable}
  \vspace{-0.1in}
\subsubsection{Results}
Figures \ref{fig:bestfit} compares our HD 33632 Ab data to the best-fitting models from the BT-Settl and Exo-REM grids.   The BT-Settl and Exo-REM best-fit models favor a similar temperature (1250--1300 $K$).  The Exo-REM models more accurately reproduce the flattened $H$ band shape, yielding a significantly lower minimum reduced $\chi^{2}$ of $\chi_{\nu}^{2}$ = 1.575 (vs. $\chi_{\nu}^{2}$ = 2.973).   The best-fit Exo-REM model favors a high surface gravity of log(g) = 5, solar metallicity, and carbon-to-oxygen ratio also near solar ([C/O] = 0.55).   The best-fit radius is 0.97 $R_{\rm Jup}$, within 10\% of values predicted for the radii of mature brown dwarfs \citep[e.g.][]{Baraffe2003}. most other Exo-REM models with $\chi_{\nu}^{2}$ $<$ 2 likewise have log(g) = 5.

\begin{figure*}[h!]
\centering
   \includegraphics[width=0.425\textwidth,trim=1mm 1mm 1mm 1mm,clip]{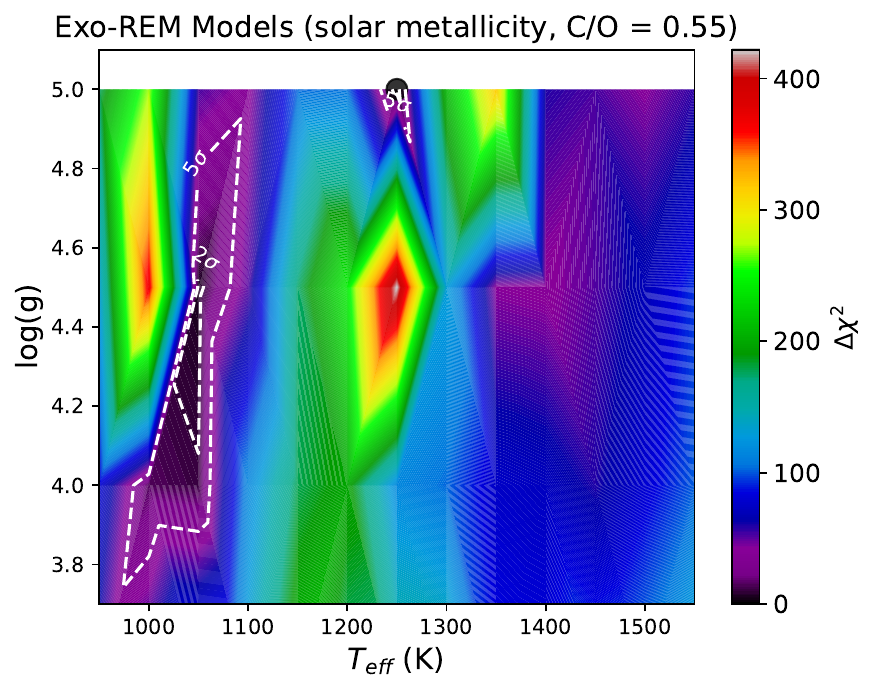}
   \includegraphics[width=0.425\textwidth,trim=1mm 1mm 1mm 1mm,clip]{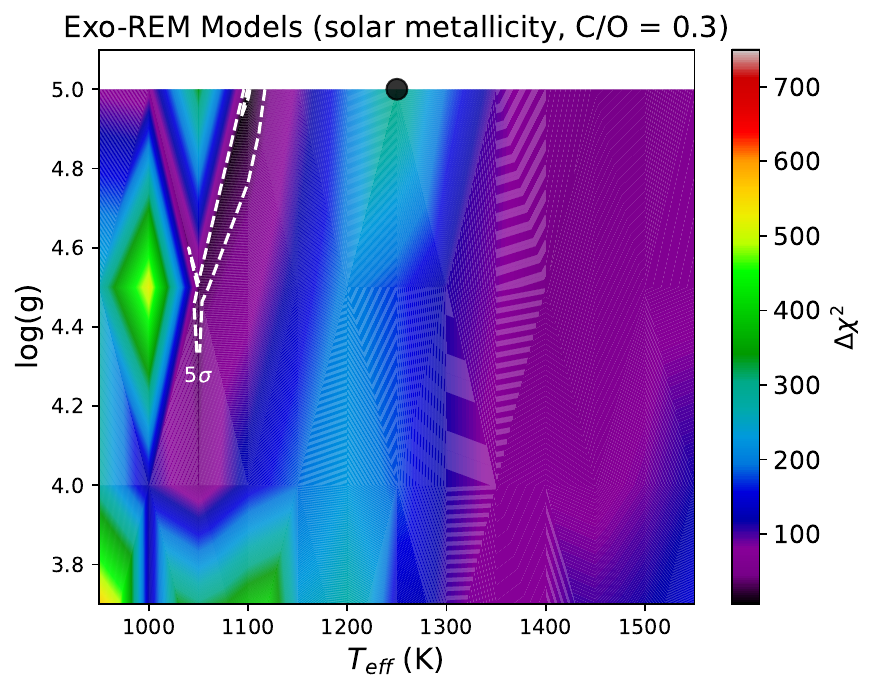}\\
   \includegraphics[width=0.425\textwidth,trim=1mm 1mm 1mm 1mm,clip]{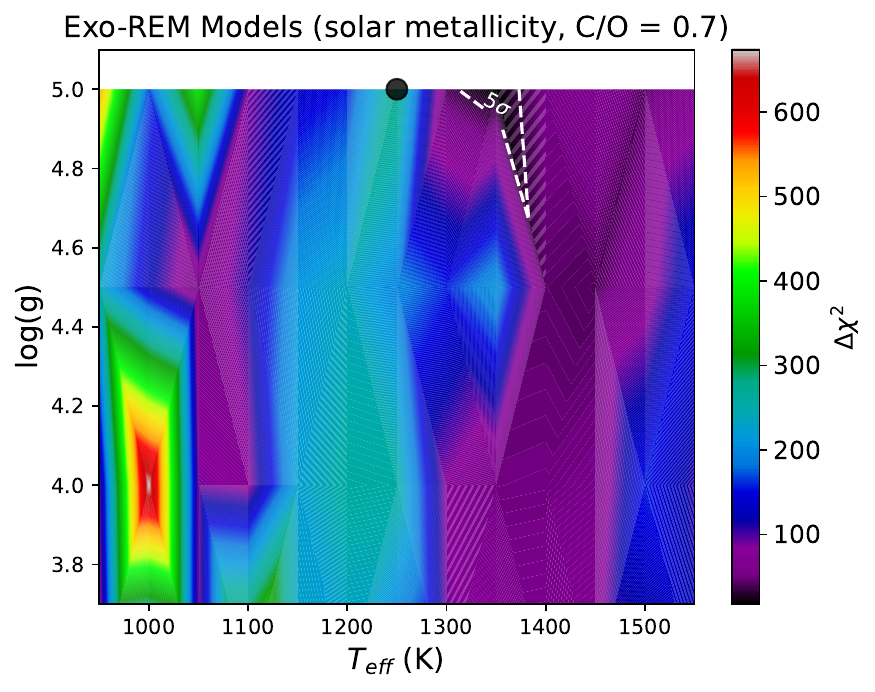}
   \includegraphics[width=0.425\textwidth,trim=1mm 1mm 1mm 1mm,clip]{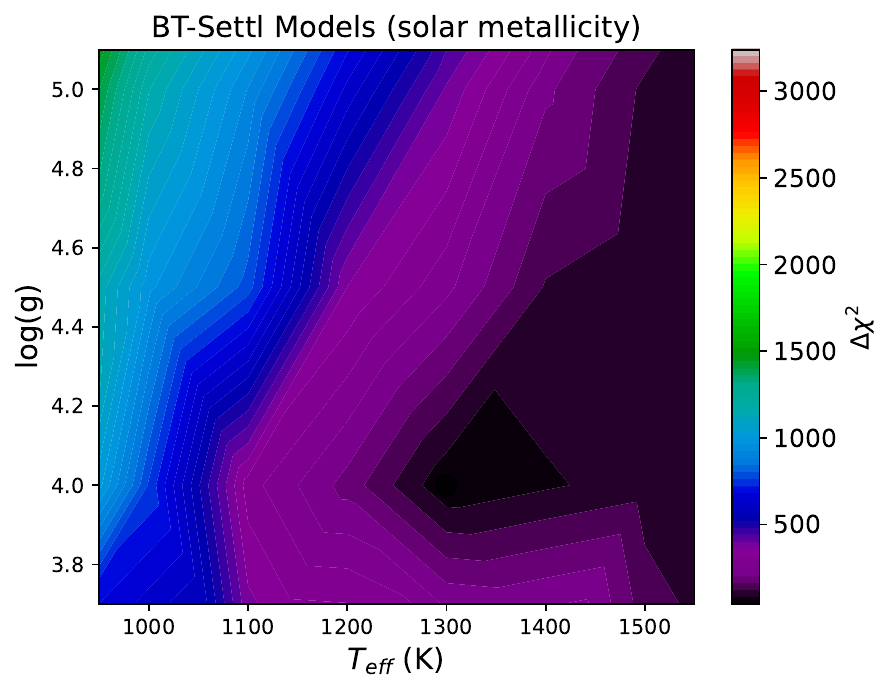}
 \vspace{-0.15in}
   \caption{Contour plots showing confidence intervals for the solar metallicity Exo-REM models at three different C/O ratios and for the BT-Settl models.  The black circle at $T_{\rm eff}$ = 1250 $K$, log(g) = 5 on the Exo-REM model panels denotes the overall best fit from the Exo-REM grid.  }
   \label{fig:bestfit_contour}
\end{figure*}
\begin{figure*}
\centering
   \includegraphics[width=0.465\textwidth,trim=1mm 1mm 1mm 1mm,clip]{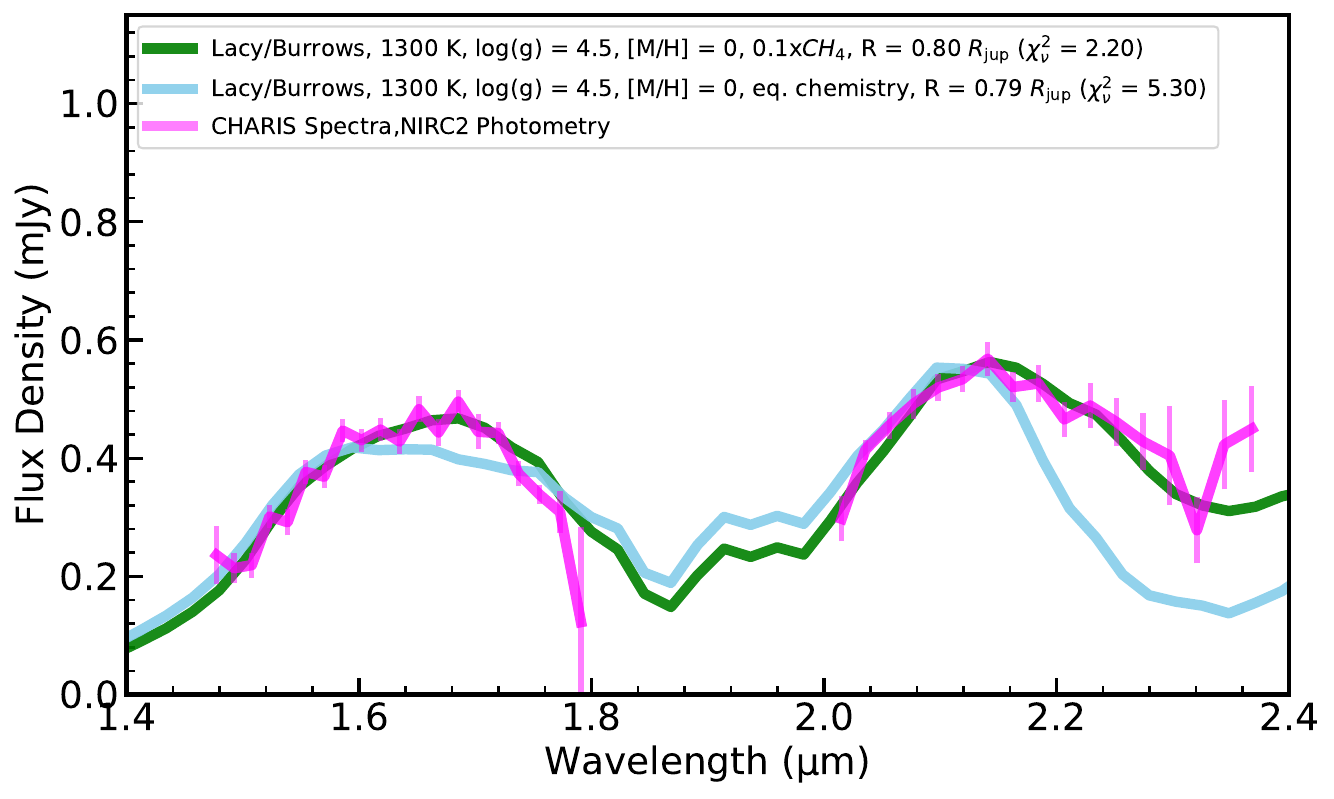}
   \includegraphics[width=0.45\textwidth,trim=1mm 1mm 1mm 1mm,clip]{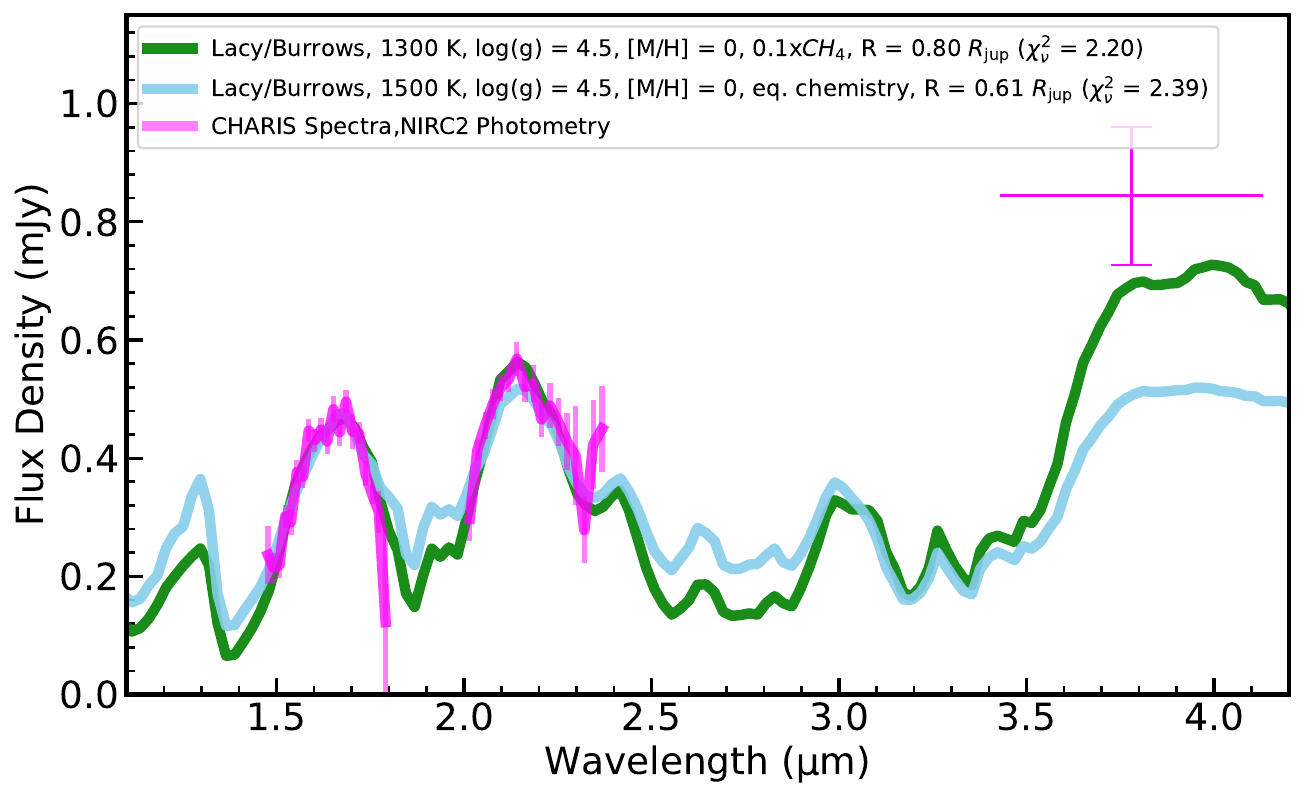}
   \caption{Same as Figure \ref{fig:bestfit} except for the best-fit Lacy/Burrows non-equilibrium chemistry and equilibrium chemistry models.}
   \label{fig:burrowsfit} 
\end{figure*}

Table \ref{exorembestfit} lists the models that match the data to within the 3-$\sigma$ confidence limit; Figure \ref{fig:bestfit_contour} shows $\Delta\chi^{2}$ contours of 2 and 5 $\sigma$ for the temperature and gravity for the Exo-REM models at solar metallicity and three different C/O ratios (0.3, 0.55, and 0.7) and the BT-Settl models.    The best-fitting Exo-REM models have temperatures of $\sim$ 1100--1350 $K$ and high gravities (log(g) $\sim$ 5) or slightly cooler temperatures (1000-1100 $K$) and lower gravities (log(g) $\sim$ 4--4.5).   Low gravity objects with temperatures of 1000-1100 $K$ are more consistent with very young ($<$ 100-200 Myr old) planet-mass objects \citep[e.g.][]{Currie2011,Barman2015,Currie2018}, while HD 33632 A is at least $\sim$ 1 Gyr old \citep[e.g.][]{Brandt2021c}.  The BT-Settl models generally fit the data poorly.


Figure \ref{fig:burrowsfit} explores the dependency of the fit on carbon chemistry using the Lacy/Burrows models.   The best-fitting model has a temperature and gravity comparable to that found from the BT-Settl and Exo-REM grids and is depleted in methane.  Equilibrium chemistry models with the same temperature and gravity predict much more methane absorption at $K$ band than seen.   While equilibrium chemistry models with higher temperatures can much better reproduce the HD 33632 Ab spectrum, their implied radii are significantly smaller, in conflict with predicted radii from cooling models.  

    \begin{figure*}[!ht]

    \begin{flushright}
    \includegraphics[width=1.0\textwidth]{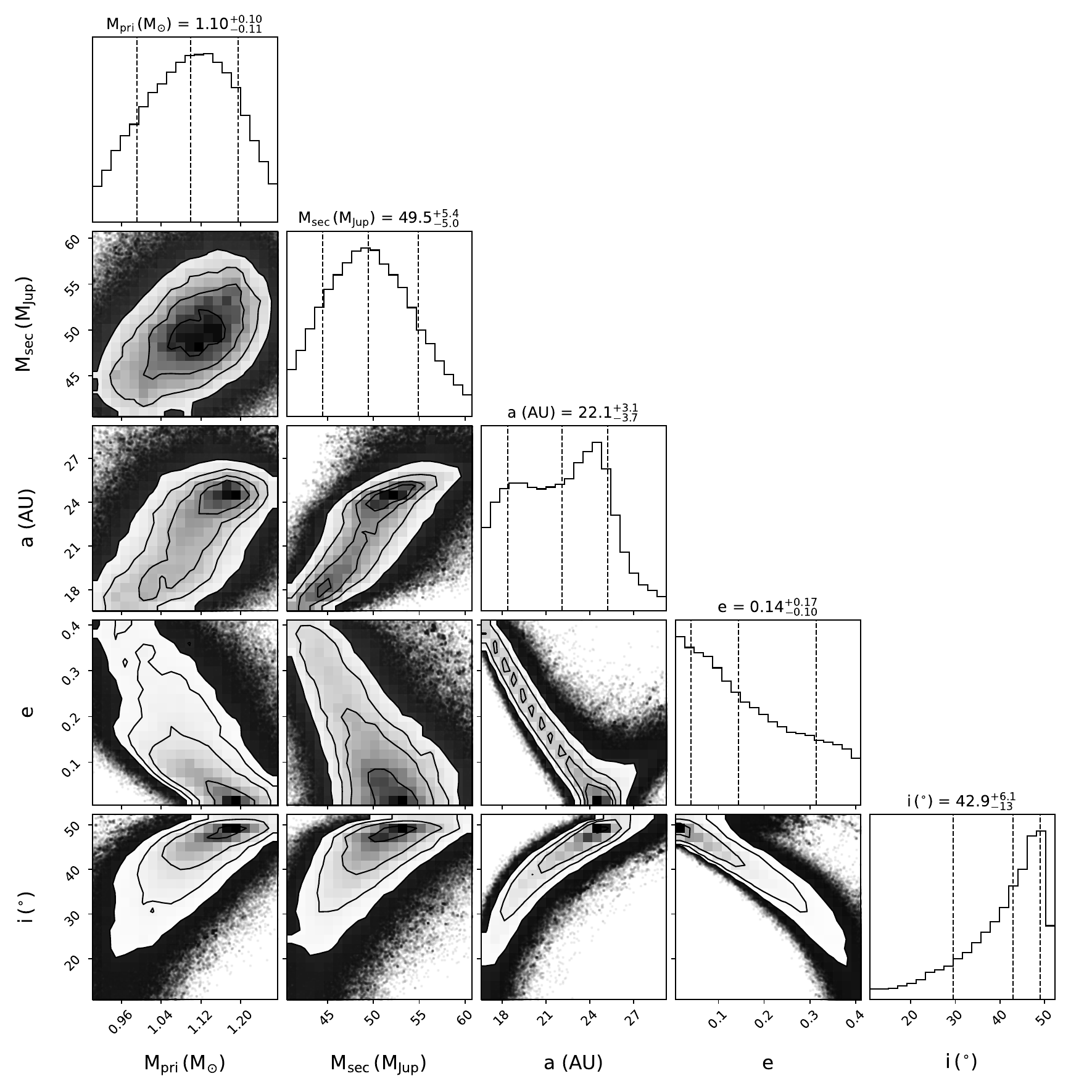} \\
    \vspace{-1.07\textwidth}
    \includegraphics[width=0.44\textwidth]{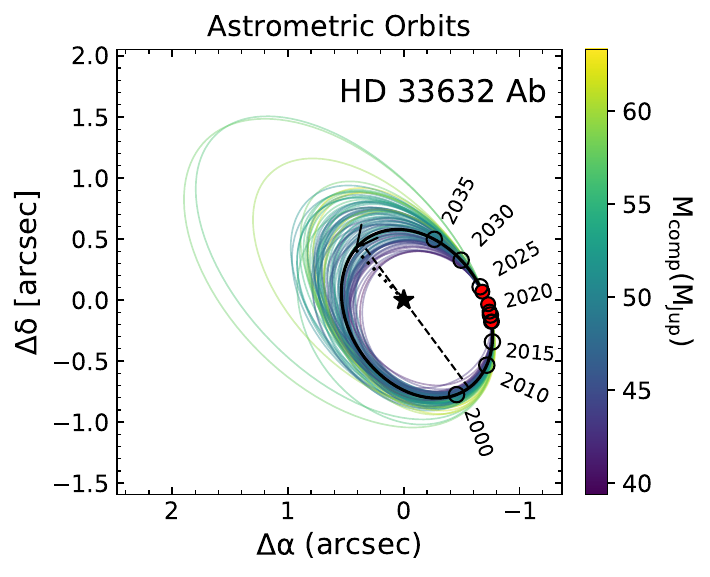} \\
    \vspace{-0.02\textwidth}
    \includegraphics[width=0.36\textwidth]{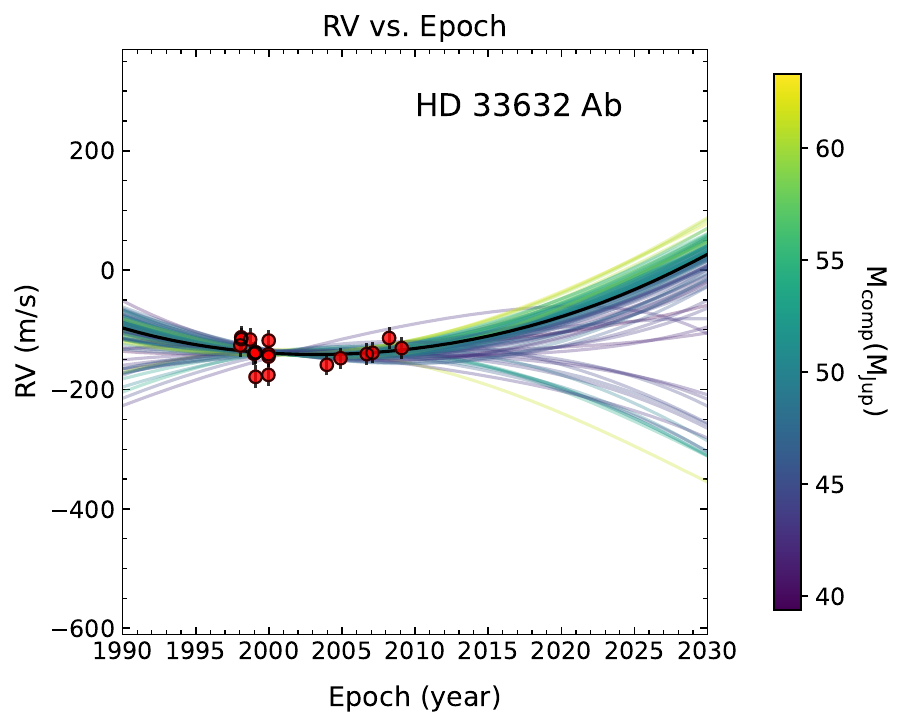}
    \vspace{0.43\textwidth}
    \end{flushright}
    \caption{Corner plot showing posterior distributions of selected orbital parameters including absolute astrometry from Hipparcos-Gaia, relative astrometry and astrometry acquired with NIRC2. The contour lines delineate the regions encompassing 68, 95, and 99\% of the posteriors.  The insets show the best-fit orbit (black curve) along with 100 orbits randomly drawn from our MCMC posterior distribution in relative astrometry (using SCExAO/CHARIS and Keck data) and radial-velocity (using only the Hamilton spectrograph data). } 
    \vspace{-0.in}
    \label{fig:corner_plot_nirc2}
\end{figure*}
   \begin{figure*}
    \begin{flushright}
        \includegraphics[width=1.0\textwidth]{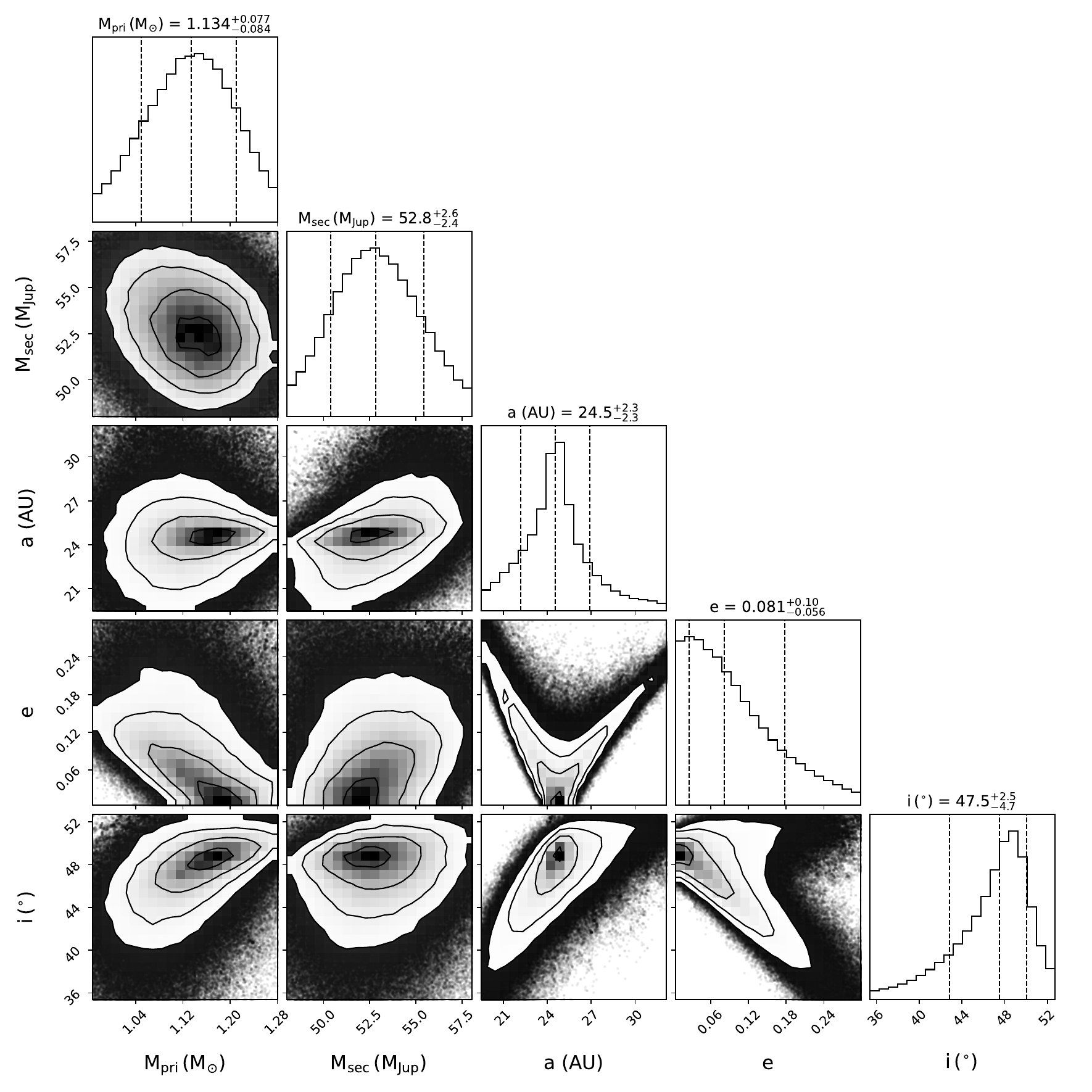} \\
    \vspace{-1.07\textwidth}
    \includegraphics[width=0.44\textwidth]{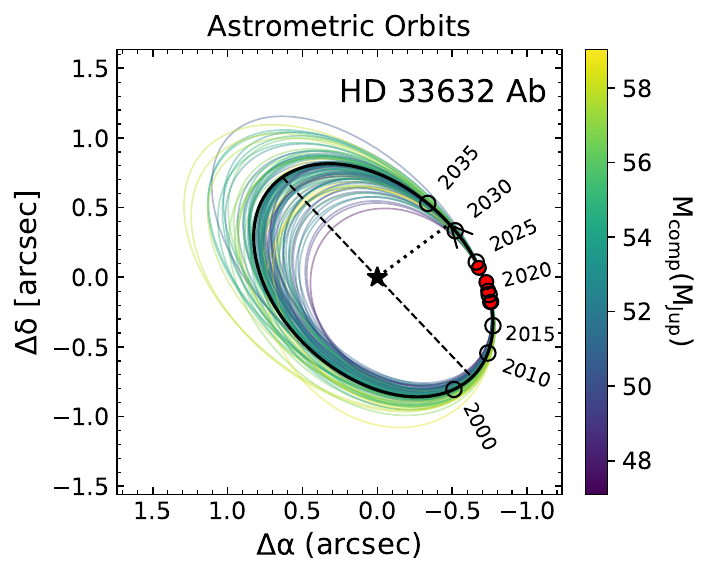} \\
    \vspace{-0.02\textwidth}
    \includegraphics[width=0.36\textwidth]{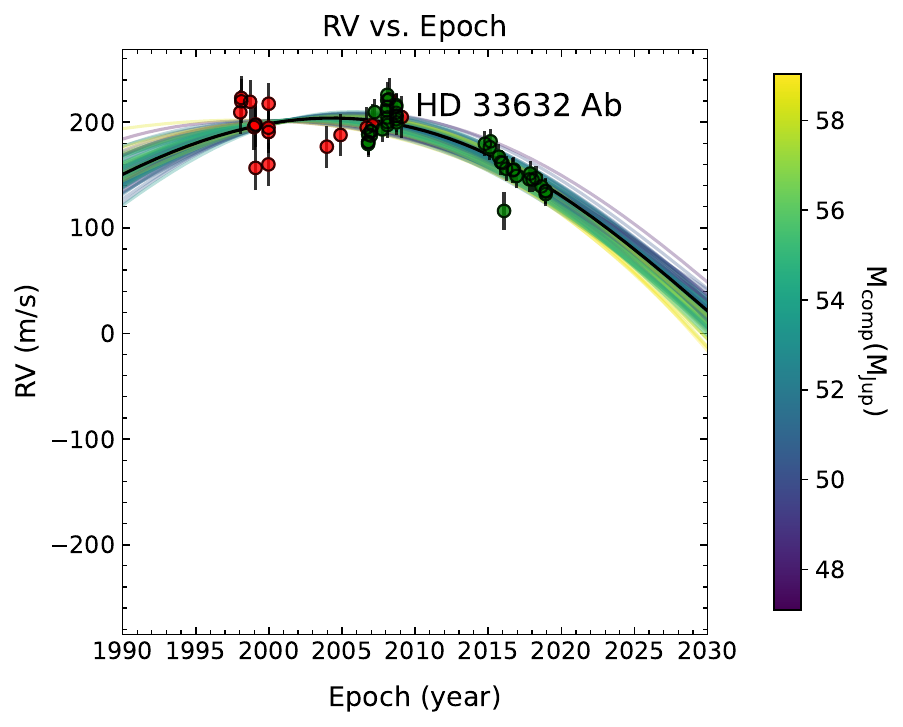}
    \vspace{0.43\textwidth}
    \end{flushright}
    \caption{Same as Figure \ref{fig:corner_plot_nirc2} but with the addition of SOPHIE RV data (green dots) added to the Hamilton spectrograph data (red dots).}
    \vspace{-0.in}
    \label{fig:corner_plot}
\end{figure*}
\noindent

\section{Dynamical Analysis} 

 We use the open-source, Python-based Markov Chain Monte Carlo (MCMC) package \texttt{orvara} \citep{Brandt2021} to fit the mass and orbit of HD33632 Ab.  To accurately measure masses and retrieve orbital parameters, \texttt{orvara} integrates a combination of radial velocities (Rvs), relative astrometry and absolute astrometry data from the Hipparcos and Gaia Catalog of Accelerations \citep[HGCA][]{Brandt2021b}. Our new data more than doubles the astrometric baseline for HD 33632 Ab and adds new RV data to better constrain HD 33632 Ab's properties.

Following previous work \citep{Currie2020a,Brandt2021c}, we adopt a prior of $1.1 \pm 0.1~M_\odot$ and absolute astrometry from Hipparcos and Gaia as reported in the HGCA using Gaia eDR3 measurements.   We include relative astrometry from SCExAO/CHARIS and NIRC2 as listed in Table \ref{astrom}.   We perform two separate fits to the data: one without the inclusion of new RV measurements from SOPHIE and the recently-published companion RV measurement from \citet{Hsu2024} (Simulation 1) and one with these data (Simulation 2)\footnote{\citet{Hsu2024} derived the absolute RV of HD 33632A\,b to be $-$8 $\pm$ 3 km s$^{-1}$ with a high-resolution spectrum of HD 33632A\,b. 
Because \texttt{orvara} employs a companion's relative RV (which is taken to be a difference between a companion's absolute RV and a host star's absolute RV), we calculated the relative RV based on \citet{Hsu2024} for HD 33632A\,b and \citet{GAIA2018} for HD 33632A.  We found that including/excluding \citet{Hsu2024} yielded a negligible difference in the companion's posterior distributions but incorporate this measurement in our simulations for completness. }.  We compare posterior distributions of the two simulations to assess the sensitivity of mass and orbital parameters derived from astrometry alone.   Because it only has a very marginal influence on the posterior distributions for HD 33632 Ab \citep{Currie2020a}, for simplicity we do not model astrometric data from the distant HD 33632 B binary.   In both cases, we fit for only one companion (HD 33632 Ab).   For both simulations, we treated the first 500 steps as burn-in and thinned the chains by a factor of 50.

For Simulation 1, we run \texttt{orvara} with 25 temperatures, 100 walkers for each temperature and 400,000 steps per walkers. The MCMC converged with a typical autocorrelation timescale of $\sim$ 45 steps per parameter.
The fitted parameters along with the priors are detailed in Table \ref{tab:mcmc_result}. The posterior distributions of the orbital parameters and the HD 33632 Ab's mass are displayed in Figure \ref{fig:corner_plot_nirc2}.

For Simulation 2 adding SOPHIE's RVs data, we run \texttt{orvara} with a total of 25 temperatures, with 100 walkers for each temperature and 500,000 steps per walkers. The MCMC converged with a typical autocorrelation timescale of $\sim$ 45 steps per parameter. 
Temperature is here is a concept in MCMC simulations with parallel tempering.  MCMC chains run at different temperatures. Hot chains (high temperatures) allow for extended exploration of the parameter space, while cold chains (lower temperatures) constrain exploration to likelihood regions. The parallel tempering approach helps avoid local minima through state swaps between hot and cold chains, enabling the convergence of the MCMC run in the case of multi-mode distributions.

We display in Table \ref{tab:mcmc_result_new} the fitted parameters and the priors of HD33632 Ab. Figure \ref{fig:corner_plot} displays posterior distributions of orbital parameters and the BD's mass for this simulation. 

Comparing the two simulations demonstrates the sensitivity of companion properties derived predominantly from astrometric data vs. those that include more extensive RV coverage.  Simulation 1's results in Figure \ref{fig:corner_plot_nirc2} largely agree with previous analyses albeit with slightly narrower posterior distributions but including some parameters (e.g. companion mass) with multiple peaks.  
We estimate a new semimajor axis, inclination, eccentricity and companion mass for HD 33632 Ab of ${22.1}_{-3.7}^{+3.1}$
au, ${42.9}_{+6.1}^{-13}$ degrees, ${0.14}_{-0.17}^{+0.10}$,  and ${49.5}_{-5.0}^{+5.4}$ $M_{\rm Jup}$, respectively.  The mass ratio is $q$ $\sim$ ${0.0432}_{+0.0043}^{-0.0051}$. 

 Compared to previous results from \citet{Brandt2021c}, the posterior distributions have comparable peaks and slightly narrower 68\% confidence intervals, indicating that additional direct imaging astrometry are not substantially improving the precision with which we extract parameters.

Adding the SOPHIE RV data signficantly improves the precision of all extracted dynamical parameters.   With these data added, HD33632 Ab's posterior distributions are consistently about half as wide.  The posterior distributions are single peaked for all parameters displayed in Figure \ref{fig:corner_plot}.   The semi-major axis is now ${24.5}_{-2.3}^{+2.3}$ au, the inclination is ${47.5}_{-4.7}^{+2.5}$ degrees, and the eccentricity is ${0.081}_{-0.056}^{+0.10}$.  HD 33632 Ab's mass and mass ratio are now constrained to within about 5\% and 10\% precision, respectively:  
 ${52.8}_{-2.4}^{+2.6}$ $M_{Jup}$ and $q$ $\sim$ ${0.0444}_{-0.0037}^{+0.0049}$.



\section{Discussion}
\subsection{Implications of Our Results For The Atmosphere  of HD 33632 Ab}
Analysis of the SCExAO/CHARIS $H$ and $K$ band spectra presented in this work further clarify HD 33632 Ab's atmospheric properties beyond the empirical comparisons to lower resolution $JHK$ CHARIS spectra presented in \citet{Currie2020a}.   HD 33632 Ab's spectra are well fit by field brown dwarfs with spectral types of L8.5--L9.5.  Atmospheric model comparisons find a best-fit temperature of 1250 $K$ and favor a high gravity (log(g)  = 5.0).   The companion's atmosphere is consistent with having a solar C/O ratio.  HD 33632 Ab has a cloudy atmosphere.  The best-fitting models (i.e. from the Exo-REM grid) incorporate disequilibrium carbon chemistry; comparisons with the Lacy \& Burrows grids show that an atmospheric model with equilibrium carbon chemistry provides poorer fits to the data, especially at $K$ band.

Formally, our results show some tension between dynamical masses and those implied from atmospheric modeling.  A log(g) = 5, 0.97 R$_{\rm Jup}$ object has a mass of $\approx$ 38 $M_{\rm Jup}$, which is 20\% lower than and inconsistent with our dynamical mass constraints.

These inconsistencies result from the limited grid phase space available for the Exo-REM models.  Holding the radius at 0.97 $R_{\rm Jup}$, a surface gravity of log(g) = 5.15 – 0.15 dex higher than our best fit – would yield perfect agreement with our dynamical mass.  Similarly, for a 1–2 Gyr-old system (see \cite{Brandt2021}), the evolutionary models from \citet{Baraffe2003} predict surface gravities of 5.10–5.26.  However, the publicly-available Exo-REM grid spans a slightly different range in surface gravity – log(g) = $3–5$ – with a coarse sampling of 0.5 dex.  In other words, we cannot explore the model phase space that would bring our dynamical masses and atmospheric model-inferred masses into agreement.  An Exo-REM grid extended to higher surface gravities may resolve this minor discrepancy.

Disequilibrium chemistry is an important contributor to consider in atmospheric models. Previous studies had shown that its effects are most significant for young, low-gravity objects near the L/T transition. However, for an older object, like HD 33632 Ab,  weaker signatures of disequilibrium chemistry are displayed, which coarsely aligns with the spectrum. 
For example, its K-band spectrum contrasts with the L/T transition object HR 8799 d, whose rising slope in F$_\nu$ space displays the absence of methane absorption.
This difference suggests that HD 33632 Ab’s atmosphere is in a chemical equilibrium. Nevertheless, comparisons with \cite{LacyBurrows2020} show that disequilibrium chemistry is still present to some extent, as equilibrium models alone overestimate the methane absorption feature.

\subsection{The Importance of RV Data for Constraining the Orbit and Dynamical Mass of HD 33632 Ab} 
Our dynamical analysis illustrates the key role that complementary RV data play even for companion search programs focused on direct imaging and astrometric data.  In contrast with earlier HD 33632 results comparing modeling results with single-epoch imaging data vs. two epochs \citep{Currie2021}, we found little improvement in mass and orbit constraints on HD 33632 Ab from additional positional measurements from imaging\footnote{It is possible that this result is specific to only a subset of brown dwarfs and planets detected from imaging and astrometry.  Analysis of other substellar companions detected with imaging and astrometry yields progressively tighter constraints on mass and orbital properties as additional imaging data are accumulated (T.C., unpublished). }.   

However, when new RV data were incorporating into our modeling, the width of the posterior distributions for HD 33632 Ab's mass and orbital properties generally shrunk by about a factor of 1.5 to 2 (e.g. a mass of ${52.8}_{-2.4}^{+2.6}$ $M_{\rm Jup}$).   RV data offer a fuller representation of companion's dynamical influence on its host star when combined with astrometry.  They should be obtained when possible for other accelerating stars that are targets of planet and brown dwarf searches, especially Sun-like stars amenable to precision RV measurements \citep[e.g. HIP 21152;][]{Kuzuhara2022,Franson2023}.

\subsection{Comparisons to Other Recent HD 33632 Ab Studies}

Our study provides a complementary probe of HD 33632 Ab's atmosphere using different data sources than \citet{Hsu2024}.  They focused on high-dispersion coronagraphic spectroscopy to detect water and carbon monoxide at peak SNRs of 4.8 and 2.7 using the cloudless Sonora Bobcat atmosphere models as molecular templates \citep{Marley2021} and SNRs of 7.8 and 6.1 using the BT-Settl models and the their baseline forward retrieval model using the \textit{petitRadTrans} code \citep{Molliere2019}, which both incorporate clouds.  Like \citet{Hsu2024}, we find evidence for disequilibrium chemistry in HD 33632 Ab's atmosphere.

Temperatures derived from their high-resolution data alone ($\sim$1550--2500 $K$) imply spectral types of M6--L4 and are easily ruled out by both our empirical comparisons and atmospheric modeling analysis.  However, their retrieval posteriors for temperature incorporating photometry and spectra from \citet{Currie2020a} are in better agreement with our results.  Orbital parameters from this work are generally consistent with previous studies \citep{Currie2020a,Brandt2021c}.   Our dynamical mass measurements without RVs are in slight tension with theirs (37$^{+7}_{-4}$ $M_{\rm Jup}$) and our masses derived with the new SOPHIE RV data rule out their results at the $\sim$1.5$\sigma$ level.  On the other hand, the usage of Gaia eDR3 data resolves the discrepancy between our parallax posterior distribution and theirs.

Our work provides a contrast to \citet{Gibbs2024} who also investigated the dynamics and atmosphere of HD 33632 Ab primarily with SCExAO/CHARIS $H$ and $K$ band spectroscopy.  Using previously-published RV data from \citet{Currie2020a}, astrometry from \citet{Currie2020a}, and a new CHARIS astrometric data point from October 2022, they estimate a dynamical mass of 51.6$^{+5.4}_{-4.8}$ $M_{\rm Jup}$: nearly identical to prior results from \citet{Brandt2021}.  Our analysis -- adding relative RV data from \citet{Hsu2024}, including new RV data for the star, and expanding HD 33632 Ab's astrometric baseline -- yields consistent results but improves the precision on HD 33632 Ab's mass, semimajor axis, eccentricity, and inclination by a factor of $\sim$2.

By fitting the CHARIS $H$ and $K$ band spectra individually to the ATMO-2020 and Sonora-Bobcat cloudless atmosphere models \citep{Phillips2020,Marley2021}, \citet{Gibbs2024} favor a temperature of $\approx$1700 $K$.  However, given HD 33632 Ab's luminosity of log$_{\rm 10}$($L/L_{\odot}$) $\sim$ -4.62, their implied radius is $\approx$0.56 $R_{\rm J}$: inconsistent with major luminosity evolution predictions of $\sim$1 $R_{\rm J}$ \citep[e.g.][]{Baraffe2003}, likely unphysical, and reminiscent of early inconsistencies between IR data for L/T transition exoplanets and atmosphere models that are insufficiently cloudy \citep[e.g.][]{Currie2011}.  They are also inconsistent with expected temperatures for L/T transition objects constrained in other studies \citep[e.g.][]{Stephens2009}.  These shortcomings are likely due to the decision to model the CHARIS $H$ and $K$ band spectra individually instead of jointly and the neglect of the NIRC2 $L_{\rm p}$ photometric data in their modeling.




\subsection{Future Observations of HD 33632 Ab}
HD 33632 Ab is a prime target for atmospheric characterization at passbands complementary to those already probed by ground-based telescopes, especially from data obtained with NASA missions.   At 3--5 $\mu m$, HD 33632 Ab has a contrast of $\approx$ 8.6 magnitudes, well within reach for detections with both NIRCam and NIRSPEC \citep{Girard2022,Ruffio2023}; HD 33632 Ab may also be detectable at 10 $\mu m$ with MIRI coronagraphy \citep{Boccaletti2024}.   These JWST modes have identified multiple molecules in the atmospheres of young exoplanets or planet-mass companions at the L/T transition (e.g. water, methane, carbon monoxide, carbon dioxide, sodium, and potassium), have constrained the presence of clouds and disquilibrium chemistry, and have more precisely determined bolometric luminosities and radii \citep[e.g.][]{Miles2023,Boccaletti2024,Franson2024}.   Similar JWST observations for HD 33632 Ab would complement these studies by exploring atmospheric properties for an older field L/T transition object with a dynamical mass measurment.

 \begin{figure}
\centering
   \includegraphics[width=0.45\textwidth,trim=0mm 55mm 0mm 60mm,clip]{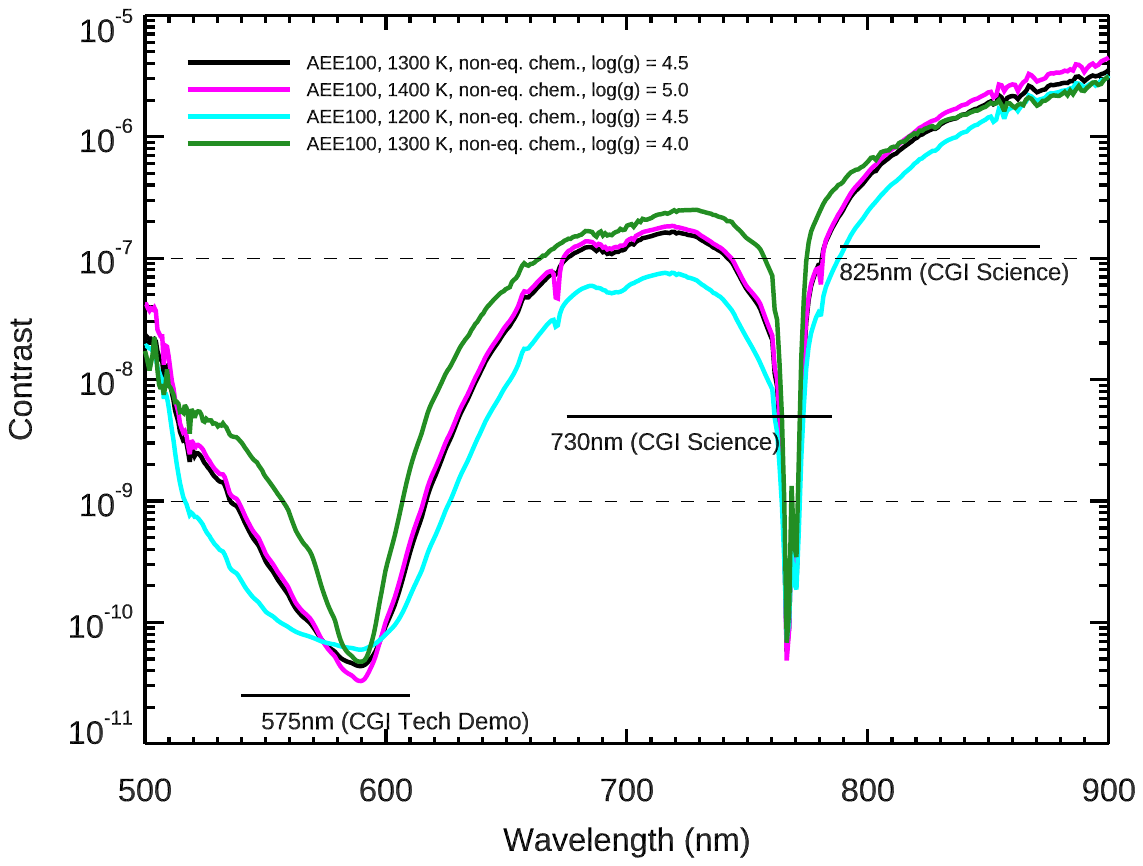}\\
   \includegraphics[width=0.45\textwidth,clip]{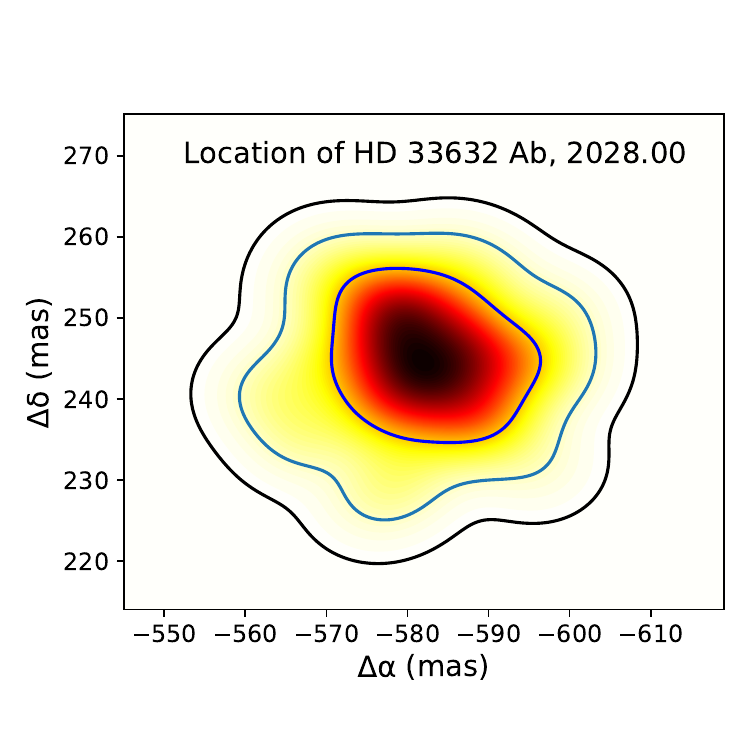}
   \caption{(Top) Predicted contrast for HD 33632 Ab in the the Roman CGI passbands given the best-fit model from the Lacy/Burrows grid (black), two other models with $\chi_{\nu}^{2}$ $\le$ 2.7 ($T_{\rm eff}$ = 1200, log(g) = 4.5; $T_{\rm eff}$ = 1400, log(g) = 5.0), and a low surface gravity model for comparison ($T_{\rm eff}$ = 1300, log(g) = 4.0).   All models assume a 100 $\mu m$ modal partical size, AEE-type clouds, and non-equilibrium carbon chemistry. (Bottom) Predicted location of HD 33632 Ab in January 2028 from the posterior distribution from \texttt{orvara} incorporating the new SOPHIE RV measurements and companion RV data from \citet{Hsu2024}. }
   \label{fig:cgidet} 
\end{figure}

To predict the detectability of HD 33632 Ab for the \textit{Nancy Grace Roman Telescope}'s Coronagraphic Instrument (CGI), we computed the companion's optical contrast for a range of Lacy \& Burrows atmosphere models centered on the best-fit model \citep[see also][]{Currie2023a}.   As shown in Figure \ref{fig:cgidet} (top panel), deep sodium absorption predicted at $\approx$ 600 nm like makes a HD 33632 Ab detection beyond the capability of CGI in the 575 nm passband to be used for CGI's technology demonstration.   For the best-fit Lacy \& Burrows mode, we predict a contrast of $\approx$ 2$\times$10$^{-10}$, while a much lower surface gravity object indicative of a young planet mass companion has weaker sodium absorption and thus may be detectable.    However, at the 730 nm passband planned for long-slit spectroscopy and 825 nm passband available for imaging, HD 33632 Ab's predicted contrast is 10$^{-7}$ and 10$^{-6}$, respectively.  The predicted location for HD 33632 early in the Roman mission (e.g. January 2028) lies beyond the 0\farcs{}45 outer working angle for suitable tech demo targets but at the edge of the dark hole region for spectroscopic mode ($\approx$ 0\farcs{}55-0\farcs{}6) and well within the dark hole region for 825 nm data (bottom panel).   Optical long-slit spectra for HD 33632 Ab cover potassium absorption, a feature whose depth and shape is expected to be sensitive to metallicity.  The full optical SED of HD 33632 Ab may also probe other atmospheric properties like clouds \citep{LacyBurrows2020}.

Finally, upcoming instruments behind ground-based extreme AO systems may similarly provide additional information on HD 33632 Ab's atmospheric properties, especially those operating at higher spectral resolution than CHARIS.   \citet{Hsu2024} demonstrated the ability of high-resolution spectroscopy to resolve HD 33632 Ab's spectral lines and constrain rotation.   Medium-resolution spectroscopy in combination with molecular mapping can yield higher SNR detections of individual molecules from applying a cross-correlation function to theoretical templates and companion spectra (e.g. \cite{Hoeijmakers2018}) without relying on a precise fiber alignment to HD 33632 Ab's position.   

The Exo-NINJA medium-resolution integral field spectrograph at Subaru \citep{ElMorsy2024} will provide this capability from the red optical through K band and will sit behind AO3K: the extreme AO upgrade to Subaru's facility AO system \citep{Lozi2022}.  Future observations with \textit{extremely-large telescopes} (e.g. \textit{Thirty Meter Telescope}) will provide even higher quality spectra. Robust atmospheric characterization of HD 33632 Ab from the optical through 10 $\mu m$ will cement it as a key object for understanding the atmospheres of substellar objects of a given mass and a reference point for understanding exoplanetary atmospheres.


\acknowledgments 
\indent The authors acknowledge the very significant cultural role and reverence that the summit of Mauna Kea holds within the Hawaiian community.  We are most fortunate to have the opportunity to conduct observations from this mountain.   We acknowledge the critical importance of the current and recent Subaru and Keck Observatory daycrew, technicians, support astronomers, telescope operators, computer support, and office staff employees and the value that the observatories have for employees and their families. The expertise, ingenuity, and dedication of Subaru and Keck staff members is indispensable to the continued successful operation of these observatories and make possible the results we present.\\
\indent The development of SCExAO was supported by JSPS (Grant-in-Aid for Research \#23340051, \#26220704 \& \#23103002), Astrobiology Center of NINS, Japan, the Mt Cuba Foundation, and the director's contingency fund at Subaru Telescope.  CHARIS was developed under the support by the Grant-in-Aid for Scientific Research on Innovative Areas \#2302.  Some of the data presented herein were obtained at the W. M. Keck Observatory, which is operated as a scientific partnership among the California Institute of Technology, the University of California and the National Aeronautics and Space Administration. The Observatory was made possible by the generous financial support of the W. M. Keck Foundation.\\
\indent T.C. is supported by NASA-Keck Strategic Mission Support grant 80NSSC24K0943.  
M.K. is supported by JSPS KAKENHI grant No. 24K07108. 
M.T. is supported by JSPS KAKENHI grant No. 24H00242.

\bibliography{bibliography}{}
\bibliographystyle{aasjournal}

\appendix
\setcounter{table}{0}
\renewcommand{\thetable}{A\arabic{table}}
\section{HD 33632 Ab CHARIS Spectra}
\begin{deluxetable}{llll}[!h]
\vspace{-.2in}
     \tablewidth{0pt}
    \tablecaption{HD 33632 Ab Spectrum}
    \tablehead{\colhead{Wavelength ($\mu$m)} & \colhead{$F_{\rm \nu}$ (mJy)} &  \colhead{$\sigma$~$F_{\rm \nu}$ (mJy)} & \colhead{SNR}}
    \startdata
1.48 & 0.236 & 0.049 & 5.0 \\
1.49 & 0.214 & 0.025 & 9.2 \\
1.51 & 0.219 & 0.021 & 10.7 \\
1.52 & 0.301 & 0.019 & 16.5 \\
1.54 & 0.292 & 0.022 & 13.6 \\
1.55 & 0.377 & 0.020 & 20.6 \\
1.57 & 0.369 & 0.020 & 19.5 \\
1.59 & 0.447 & 0.020 & 24.2 \\
1.6 & 0.431 & 0.019 & 24.0 \\
1.62 & 0.448 & 0.020 & 24.1 \\
1.63 & 0.428 & 0.021 & 21.5 \\
1.65 & 0.483 & 0.022 & 23.3 \\
1.67 & 0.444 & 0.023 & 20.5 \\
1.69 & 0.496 & 0.019 & 28.3 \\
1.7 & 0.445 & 0.029 & 15.8 \\
1.72 & 0.442 & 0.019 & 24.5 \\
1.74 & 0.374 & 0.021 & 18.5 \\
1.76 & 0.339 & 0.017 & 21.8 \\
1.77 & 0.309 & 0.035 & 11.3 \\
1.79 & 0.123 & 0.160 & 0.8 \\
2.02 & 0.297 & 0.038 & 9.3 \\
2.04 & 0.414 & 0.017 & 27.7 \\
2.06 & 0.455 & 0.023 & 23.2 \\
2.08 & 0.492 & 0.025 & 22.9 \\
2.1 & 0.519 & 0.023 & 25.0 \\
2.12 & 0.534 & 0.022 & 29.5 \\
2.14 & 0.568 & 0.029 & 21.6 \\
2.16 & 0.52 & 0.025 & 23.5 \\
2.18 & 0.526 & 0.032 & 18.0 \\
2.21 & 0.465 & 0.030 & 17.0 \\
2.23 & 0.489 & 0.038 & 13.7 \\
2.25 & 0.461 & 0.040 & 12.2 \\
2.27 & 0.428 & 0.049 & 9.2 \\
2.3 & 0.404 & 0.084 & 5.0 \\
2.32 & 0.278 & 0.056 & 5.1 \\
2.34 & 0.423 & 0.076 & 5.8 \\
2.37 & 0.45 & 0.073 & 6.9 \\
\enddata
\tablecomments{Throughput of HD 33632 Ab spectrum in H and K bands extracted from 18 October 2021 and reduced by A-LOCI. }
\label{tab:spectrum_snr}
\end{deluxetable}

\newpage
\section{New SOPHIE Radial-Velocity Measurements}

\begin{deluxetable}{ccc}[ht]
\label{tab:rvs_sophie}
     \tablewidth{0pt}
    \tablecaption{RV data from SOPHIE}
    \tablehead{\colhead{MBJD} & \colhead{RV (m/s)} &  \colhead{RV$_{err}$ (m/s)}}
    \startdata
54023.135 & 1.646 & 2.533 \\
54023.145 & -0.191 & 2.418 \\
54024.098 & 8.936 & 2.348 \\
54026.138 & 1.903 & 2.553 \\
54086.061 & 12.004 & 3.107 \\
54089.965 & 7.944 & 2.286 \\
54096.954 & 11.959 & 2.105 \\
54186.829 & 30.134 & 3.002 \\
54375.167 & 13.770 & 2.262 \\
54496.821 & 31.587 & 2.228 \\
54500.781 & 41.415 & 2.282 \\
54501.797 & 34.976 & 2.601 \\
54502.813 & 46.418 & 2.468 \\
54504.802 & 17.816 & 2.525 \\
54505.812 & 20.640 & 2.281 \\
54506.821 & 25.070 & 2.219 \\
54718.146 & 29.432 & 2.530 \\
54724.106 & 36.658 & 2.807 \\
54725.156 & 35.151 & 2.437 \\
54726.167 & 23.320 & 2.158 \\
54729.153 & 26.512 & 2.073 \\
54733.165 & 20.353 & 2.397 \\
54734.136 & 26.740 & 2.429 \\
56939.063 & 0.130 & 1.371 \\
57063.834 & -3.151 & 1.472 \\
57080.880 & 2.432 & 1.537 \\
57289.088 & -12.200 & 1.581 \\
57342.111 & -17.254 & 1.507 \\
57378.067 & -18.113 & 1.480 \\
57416.955 & -63.459 & 13.328 \\
57468.849 & -23.497 & 1.300 \\
57637.133 & -25.059 & 1.417 \\
57664.165 & -24.301 & 1.856 \\
57729.130 & -30.320 & 0.947 \\
58051.112 & -33.556 & 1.524 \\
58073.015 & -28.097 & 1.372 \\
58130.871 & -33.939 & 1.166 \\
58214.805 & -32.456 & 1.532 \\
58351.140 & -40.216 & 2.098 \\
58457.026 & -45.129 & 1.359 \\
58457.944 & -47.507 & 1.413 \\
58459.023 & -44.478 & 1.360 \\
\enddata
\end{deluxetable}

\newpage
\section{Tables of \texttt{orvara} Modeling Results}

\begin{deluxetable*}{lccr}[htb!]
\tablecaption{MCMC Orbit Fitting Priors and Results}
\tablewidth{0pt}
  \tablehead{
  \colhead{Parameter} & \colhead{16/50/84\% quantiles} & Prior}
  \startdata
\multicolumn{3}{c}{Fitted Parameters} \\ \hline
RV jitter (m/s)  &      ${18.2}_{-3.3}^{+4.4}$ & log-flat \\
$M_{\rm pri}$ ($M_\odot$)   &      ${1.10}_{-0.11}^{+0.10}$  & Gaussian, $1.1 \pm 0.1$ \\
$M_{\rm sec}$ ($M_{\rm Jup}$)  &       ${49.5}_{-5.0}^{+5.4}$  & $1/M_{\rm sec}$ (log-flat) \\
Semimajor axis $a$ (au)    &     ${22.1}_{-3.7}^{+3.1}$ & $1/a$  (log-flat) \\
$\sqrt{e} \sin \omega$\tablenotemark{\rm *}    &     ${-0.01}_{-0.18}^{+0.16}$ & uniform \\
$\sqrt{e} \cos \omega$\tablenotemark{\rm *}  &       ${0.23}_{-0.45}^{+0.27}$  & uniform  \\
Inclination ($^\circ$)    &     ${42.9}_{-13}^{+6.1}$   &  $\sin i$ (geometric) \\
PA of the ascending node $\Omega$ ($^\circ$)  &      ${41.4}_{-5.3}^{+145}$ & uniform \\
Mean longitude at 2010.0 ($^\circ$) &        ${201}_{-22}^{+10}$ & uniform \\
Parallax (mas)   &      ${37.8953}_{-0.0067}^{+0.0067}$  & Gaussian, 0 \\
\hline
\multicolumn{3}{c}{Derived Parameters} \\ \hline
Period (yrs)     &    ${97}_{-21}^{+19}$ \\
Argument of periastron $\omega$ ($^\circ$)\tablenotemark{\rm *}   &      ${189}_{-175}^{+150}$ \\
Eccentricity $e$    &   ${0.14}_{-0.10}^{+0.17}$   \\
Semimajor axis (mas)  &   ${838}_{-142}^{+118}$ \\
Periastron time $T_0$ (JD)    &    ${2468053}_{-2531}^{+4026}$\\
Mass ratio   &      ${0.0432}_{-0.0043}^{+0.0051}$
\enddata
\tablenotetext{*}{Orbital parameters listed are of HD 33632 Ab.  HD 33632 Aa has $\omega$ shifted by 180$^\circ$. }
\label{tab:mcmc_result}
\end{deluxetable*}

\begin{deluxetable*}{lccr}[htb!]
\tablecaption{MCMC Orbit Fitting Priors and Results (including RV Data from SOPHIE and Companion RV Data from \citealt{Hsu2024})}
\tablewidth{0pt}
  \tablehead{
    \colhead{Parameter} &
    \colhead{16/50/84\% quantiles} &
    \colhead{Prior}}
  \startdata
\multicolumn{3}{c}{Fitted Parameters} \\ \hline
RV jitter (m/s)  &       ${20.5}_{-3.5}^{+4.5}$, ${11.6}_{-1.3}^{+1.6}$ & log-flat \\
$M_{\rm pri}$ ($M_\odot$)   &      ${1.134}_{-0.084}^{+0.077}$  & Gaussian, $1.1 \pm 0.1$ \\
$M_{\rm sec}$ ($M_{\rm Jup}$)  &      ${52.8}_{-2.4}^{+2.6}$ & $1/M_{\rm sec}$ (log-flat) \\
Semimajor axis $a$ (AU)    &     ${24.5}_{-2.3}^{+2.3}$ & $1/a$  (log-flat) \\
$\sqrt{e} \sin \omega$\tablenotemark{\rm *}    &     ${0.09}_{-0.18}^{+0.16}$ & uniform \\
$\sqrt{e} \cos \omega$\tablenotemark{\rm *}  &       ${-0.09}_{-0.26}^{+0.28}$ & uniform  \\
Inclination ($^\circ$)    &     ${47.5}_{-4.7}^{+2.5}$  &  $\sin i$ (geometric) \\
PA of the ascending node $\Omega$ ($^\circ$)  &      ${220.8}_{-3.3}^{+2.3}$ & uniform \\
Mean longitude at 2010.0 ($^\circ$) &        ${22.8}_{-5.2}^{+7.3}$ & uniform \\
Parallax (mas)   &      ${37.8954}_{-0.0067}^{+0.0067}$ & Gaussian, 0 \\
\hline
\multicolumn{3}{c}{Derived Parameters} \\ \hline
Period (yrs)     &    ${111}_{-14}^{+19}$ \\
Argument of periastron $\omega$ ($^\circ$)\tablenotemark{\rm *}   &      ${144}_{-98}^{+58}$\\
Eccentricity $e$    &   ${0.081}_{-0.056}^{+0.10}$  \\
Semimajor axis (mas)  &   ${930}_{-89}^{+89}$\\
Periastron time $T_0$ (JD)    &     ${2468247}_{-9010}^{+6088}$ \\
Mass ratio   &     ${0.0444}_{-0.0037}^{+0.0049}$
\enddata
\tablenotetext{*}{Orbital parameters listed are of HD 33632 Ab.  HD 33632 Aa has $\omega$ shifted by 180$^\circ$. }
\label{tab:mcmc_result_new}
\end{deluxetable*}

\end{document}